\documentclass[fleqn,10pt]{article}
\usepackage{latexsym, graphicx, epsfig, amsmath, notation, amssymb,amsfonts}
\usepackage{natbib,amsthm,version}
\usepackage{amsbsy,bm,multirow}
\usepackage[titletoc,page]{appendix}
\usepackage[mathscr]{eucal}
\usepackage{enumerate}

\let\emptyset\varnothing

\title{
A Pressure Correction Scheme for  Generalized
Form of Energy-Stable Open Boundary Conditions
for Incompressible Flows
} 
\author{
  S. Dong\thanks{Author of correspondence. Email: sdong@purdue.edu}, \
   J. Shen \\
  Center for Computational \& Applied Mathematics \\
  Department of Mathematics \\
  Purdue University 
 } 

\date{} 
\begin{document}
\maketitle



\begin{abstract}

We present a generalized form of open boundary 
conditions, and an associated numerical algorithm,
for simulating incompressible flows
involving open or outflow boundaries.
The generalized form represents a family of open
boundary conditions, which all ensure the
energy stability of the system,
even in situations where  strong vortices 
or backflows occur at the 
open/outflow boundaries.
Our numerical algorithm for 
treating these open boundary conditions 
is based on a rotational
pressure correction-type strategy,
with a formulation suitable for
 $C^0$ spectral-element spatial discretizations.
We have introduced a discrete equation and associated
boundary conditions for an auxiliary variable.
The algorithm contains constructions that prevent 
a numerical locking at the open/outflow boundary.
In addition, we have also developed a scheme 
with a provable unconditional stability
for a sub-class of the open boundary conditions.
Extensive numerical experiments have been presented
to demonstrate the performance of our method 
for several flow problems involving 
open/outflow boundaries.
We compare simulation results with the 
experimental data to demonstrate the accuracy 
of our algorithm.
Long-time simulations have been performed for a range of Reynolds 
numbers at which strong vortices or backflows occur
at the open/outflow boundaries. 
We show that
the open boundary conditions and the numerical
algorithm developed herein produce stable simulations
in such situations.

\end{abstract}


\vspace{0.05cm}
Keywords: {\em 
pressure correction scheme;
outflow; open boundary condition; outflow boundary condition;
 unbounded domain; backflow instability;
spectral element
}

\section{Introduction}
\label{sec:intro}

%
%
%


Outflows or open boundaries are a crucial issue to 
incompressible flow simulations. 
Many types of flows, such as wakes, jets and shear layers,
involve physically unbounded domains.
To numerically simulate such problems, it is necessary
to artificially truncate the domain to finite sizes.
Therefore, some open  boundary condition (OBC)
will be required at the artificial boundary \cite{Tsynkov1998}.
Open boundary conditions are also referred to 
as outflow  boundary conditions or
artificial boundary conditions
in the literature.
These boundary conditions have been under
intensive studies by the community for
decades, and a  
large volume of work has been accumulated.
Some of the desirable features of an
ideal method are summarized in 
e.g. \cite{SaniG1994}. 
A review of the status of the field up to 
the mid-1990s can be found in \cite{SaniG1994,Gresho1991};
see also the references therein.
Among the existing techniques
the traction-free boundary condition or
its variants (e.g. no-flux condition) 
\cite{TaylorRM1985,Gartling1990,EngelmanJ1990,
Leone1990,BehrLST1991,SaniG1994,GuermondMS2005,Liu2009}
and the convective boundary condition
\cite{Orlanski1976,Gresho1991,
KeskarL1999,OlshanskiiS2000,ForestierPPS2000,RuithCM2004,CraskeR2013}
are some of the most commonly used.
A variety of other methods have also 
been contributed by various researchers;
see e.g. \cite{PapanastasiouME1992,JinB1993,Johansson1993,HeywoodRT1996,
Griffiths1997,Renardy1997,FormaggiaGNQ2002,
HasanAS2005,NordstromS2005,NordstromMS2007,
GrinbergK2008,KimFHJT2009,PouxGA2011,PouxGAA2012},
among others.


A commonly-encountered issue with 
outflows is the numerical instability associated 
with strong vortices or backflows at the 
open/outflow boundaries.
It is often referred to as the backflow instability.
When strong vortices or backflows occur 
at the open boundaries,
the computation is observed to  instantly become unstable 
(see e.g. \cite{DongK2005,DongKER2006},
among others).
If the Reynolds number is low, the presence of
a certain amount of backflow or vortices at 
the open/outflow boundary usually does not cause 
difficulty. 
But when the Reynolds number increases beyond
some moderate value,
typically about several hundred to a thousand
depending on the flow geometry,
this numerical instability becomes a severe issue
for simulations.
It is observed that 
reducing the time step size or increasing the grid
resolution does not help with this instability.

In production simulations 
a usual remedy for this problem is to employ
a large computational domain for a given Reynolds number to be
simulated, 
such that
the outflow boundary can be placed far downstream
and sufficiently away from the region of 
interest \cite{DongKER2006}.
As a result, the vortices generated in the region of interest
can be sufficiently dissipated before reaching
the outflow boundary
for the given Reynolds number. 
At high Reynolds numbers, the domain size
essential for numerical stability 
can become very substantial \cite{DongKER2006}.
%
As pointed out by \cite{DongKC2013},
the drawback here is that the large computational domain
requires larger meshes and induces 
increased computational costs.
In addition, this strategy is not scalable
 with respect to the Reynolds number,
because the domain size essential for numerical stability
grows with increasing Reynolds number.


In the literature there exist several open boundary conditions
that are effective for coping with
the backflow instability.
The earliest one appears to be from \cite{BruneauF1994};
see also \cite{BruneauF1996}.
Based on a symmetrization of the nonlinear term and
the weak form of the incompressible Navier-Stokes equation,
a modified traction condition containing a term with the form 
($\mathbf{n}$ denoting the directional vector at boundary
and $\mathbf{u}$ denoting the velocity)
\begin{equation}
\frac{1}{2}(\mathbf{n}\cdot\mathbf{u})^{-}\mathbf{u}, \quad
\text{where} \
(\mathbf{n}\cdot\mathbf{u})^{-} = \left\{
\begin{array}{ll}
\mathbf{n}\cdot\mathbf{u}, & \text{if} \ \mathbf{n}\cdot\mathbf{u}<0 \\
0, & \text{otherwise}
\end{array}
\right.
\label{equ:obc_term_form}
\end{equation}
was imposed on the outflow boundary in \cite{BruneauF1994}.
Note that the original form given in \cite{BruneauF1994}
includes a base velocity profile
that is assumed to be known.
This form of the open boundary condition has also appeared 
in later works by other researchers; see e.g. \cite{LanzendorferS2011}
among others.
%
In \cite{BazilevsGHMZ2009,Moghadametal2011,GravemeierCYIW2012,IsmailGCW2014},
the traction in the open boundary condition 
contains a term 
with a similar form, $(\mathbf{n}\cdot\mathbf{u})^{-}\mathbf{u}$,
but without the $\frac{1}{2}$ factor compared to \cite{BruneauF1994}. 
Note that in \cite{BazilevsGHMZ2009}
the boundary conditions are given separately in
the normal and tangential directions,
and in \cite{Moghadametal2011} a form 
$\beta(\mathbf{n}\cdot\mathbf{u})^{-}\mathbf{u}$,
where $0< \beta < 1$ is a constant,
has also been considered.
%
Based on the energy
balance relation of the system,
an open boundary condition is recently proposed
in \cite{DongKC2013}, which contains
a term of the form,
$
\frac{1}{2}\left|\mathbf{u} \right|^2\mathbf{n}\Theta_0(\mathbf{n}\cdot\mathbf{u})
$,
where $\Theta_0(\mathbf{n}\cdot\mathbf{u})$ is 
a smoothed step function about $\mathbf{n}\cdot\mathbf{u}$
(see also Section \ref{sec:obc}),
and $|\mathbf{u}|$ denotes the magnitude of the
velocity.
While the function $\Theta_0(\mathbf{n}\cdot\mathbf{u})$
plays a role comparable to that of 
$(\mathbf{n}\cdot\mathbf{u})^-$ defined in \eqref{equ:obc_term_form}, 
the form $\frac{1}{2}\left|\mathbf{u} \right|^2\mathbf{n}$
from \cite{DongKC2013}
is very different from  those involving
$(\mathbf{n}\cdot\mathbf{u})\mathbf{u}$
by the other 
researchers \cite{BruneauF1994,BazilevsGHMZ2009,Moghadametal2011,GravemeierCYIW2012,IsmailGCW2014}.
%
Another open boundary condition is proposed
in a recent study \cite{BertoglioC2014},
in which the tangential velocity derivative 
at the open boundary
is penalized to allow for an improved
energy balance.


In the current paper, we present a generalized form of the
open boundary conditions that ensure 
the energy stability of the system.
The generalized form represents a family of open
boundary conditions.
It contains 
the open boundary conditions
of \cite{BruneauF1994,BazilevsGHMZ2009,GravemeierCYIW2012,IsmailGCW2014,DongKC2013} 
as particular cases.
In addition, it also provides new forms 
of energy-stable open boundary conditions.
We further present an algorithm
for numerically treating 
the generalized open boundary conditions
based on a pressure correction-type strategy.
It is noted that in \cite{DongKC2013}
a splitting scheme based on a rotational
velocity correction-type strategy \cite{GuermondS2003a,DongS2010} has been
developed for dealing with the proposed open boundary
condition therein.
The numerical algorithm developed in
the current work is based on a different 
strategy, and has a different 
algorithmic formulation.
We refer to \cite{GuermondMS2006}
and the references therein
for a review of the pressure-correction
idea and an exposition of related
concepts.
The main algorithm in the current paper 
is semi-implicit and
conditionally stable in nature.
In addition, we also present
a rotational pressure correction scheme
with a provable unconditional stability
for a sub-class of the generlized
open boundary conditions.


The novelties of this paper lie in three
aspects: 
(i) the generalized form of energy-stable
open boundary conditions,
(ii) the rotational pressure correction-type algorithm
for treating the proposed
open boundary conditions, and
(iii) the unconditionally stable scheme for
a sub-class of the open boundary conditions.
%
%


We employ $C^0$ spectral 
elements \cite{SherwinK1995,KarniadakisS2005,ZhengD2011}
for spatial discretizations in
the current paper.
The algorithmic formulation presented here
without change 
 also applies to low-order
finite elements.
It should be noted that
the open boundary conditions and the numerical 
algorithm for treating these boundary 
conditions developed herein 
are general, and can also be used 
with other spatial discretizations
such as finite difference and finite volume.



\section{Open Boundary Conditions and Algorithm}
\label{sec:method}



\subsection{A Generalized Form of Open Boundary Conditions}
\label{sec:obc}


Let $\Omega$ denote the flow domain in two or three dimensions (2-D or 3-D),
and $\partial\Omega$ denote the domain boundary.
We consider the incompressible flow contained within $\Omega$,
which is described by the normalized incompressible Navier-Stokes
equations:
\begin{subequations}
\begin{align}
&
\frac{\partial\mathbf{u}}{\partial t} + \mathbf{u}\cdot\nabla\mathbf{u}
  = -\nabla p + \nu\nabla^2\mathbf{u} + \mathbf{f},
\label{equ:nse} \\
&
\nabla\cdot\mathbf{u} = 0,
\label{equ:continuity}
\end{align}
\end{subequations}
where $\mathbf{u}(\mathbf{x},t)$ is the velocity, $p(\mathbf{x},t)$
is pressure, $\mathbf{f}(\mathbf{x},t)$ is an external body force,
and $\mathbf{x}$ and $t$ are respectively the spatial coordinate
and time.
$\nu=\frac{1}{Re}$ is the non-dimensional fluid viscosity, and
$Re$ is the Reynolds number defined after appropriately choosing
a characteristic velocity scale and a characteristic length scale.


Let us assume that 
$
\partial\Omega = \partial\Omega_d \cup \partial\Omega_o,
$
and $\partial\Omega_d \cap \partial\Omega_o = \emptyset$.
$\partial\Omega_d$ is the Dirichlet boundary, on which
the velocity is known,
\begin{equation}
\mathbf{u} = \mathbf{w}(\mathbf{x},t),
\quad\quad
\text{on } \partial\Omega_d,
\label{equ:bc_dbc}
\end{equation}
where $\mathbf{w}$ is the boundary velocity.
On $\partial\Omega_o$ neither the velocity $\mathbf{u}$ nor 
the pressure $p$ is known.
We will refer to $\partial\Omega_o$ as 
the open (or outflow) boundary hereafter in the paper.


We consider the following boundary conditions for
the open boundary $\partial\Omega_o$,
\begin{equation}
\text{\underline{OBC-A}:} \
-p\mathbf{n} + \nu\mathbf{n}\cdot\nabla\mathbf{u}
  - \frac{1}{4}\left[\left|\mathbf{u} \right|^2 \mathbf{n}  
      +  \left(\mathbf{n} \cdot \mathbf{u} \right)\mathbf{u}
    \right] \Theta_0(\mathbf{n},\mathbf{u})
= \mathbf{f}_b(\mathbf{x},t),
\quad\quad
\text{on } \partial\Omega_o,
\label{equ:bc_obc_A}
\end{equation}
\begin{equation}
\text{\underline{OBC-B}:} \
-p\mathbf{n} + \nu\mathbf{n}\cdot\nabla\mathbf{u}
  - \left(\left|\mathbf{u} \right|^2 \mathbf{n}  \right)     
    \Theta_0(\mathbf{n},\mathbf{u})
= \mathbf{f}_b(\mathbf{x},t),
\quad\quad
\text{on } \partial\Omega_o,
\label{equ:bc_obc_B}
\end{equation}
\begin{equation}
\text{\underline{OBC-C}:} \
-p\mathbf{n} + \nu\mathbf{n}\cdot\nabla\mathbf{u}
  - \frac{1}{2}\left[\left|\mathbf{u} \right|^2 \mathbf{n}  
      +  \left(\mathbf{n} \cdot \mathbf{u} \right)\mathbf{u}
    \right] \Theta_0(\mathbf{n},\mathbf{u})
= \mathbf{f}_b(\mathbf{x},t),
\quad\quad
\text{on } \partial\Omega_o,
\label{equ:bc_obc_D}
\end{equation}
where $\mathbf{n}$ is the outward-pointing unit vector normal
to $\partial\Omega_o$, and $\left|\mathbf{u} \right|$ is the magnitude
of the velocity $\mathbf{u}$.
$\mathbf{f}_b$ is a function on $\partial\Omega_o$
for the purpose of numerical testing only, and will be set to
$f_b=0$ in actual simulations.
$\Theta_0$ is a smoothed step function  given by
\begin{equation}
\Theta_0(\mathbf{n},\mathbf{u}) = \frac{1}{2}\left(
    1 - \tanh \frac{\mathbf{n}\cdot \mathbf{u}}{\delta U_0}
  \right)
\label{equ:Theta_0_expr}
\end{equation}
where $U_0$ is the characteristic velocity scale, and
$\delta$ is a non-dimensional positive constant that is sufficiently small.
The parameter $\delta$ controls the sharpness of the
smoothed step function, and it is sharper if $\delta$ is smaller.
As $\delta \rightarrow 0$, $\Theta_0(\mathbf{n},\mathbf{u})$
approaches the step function.
When $\delta$ is sufficiently small,
$\Theta_0(\mathbf{n},\mathbf{u})$ takes essentially 
the unit value where $\mathbf{n}\cdot\mathbf{u}<0$
and vanishes otherwise.
The simulation result is not sensitive
to $\delta$ when it is sufficiently small; see \cite{DongKC2013}.

In addition, we also consider
the following conditions:
\begin{equation}
\text{\underline{OBC-D}:} \
-p\mathbf{n} + \nu\mathbf{n}\cdot\nabla\mathbf{u}
  - \left[ 
       \left(\mathbf{n} \cdot \mathbf{u} \right)\mathbf{u}
    \right] \Theta_0(\mathbf{n},\mathbf{u})
= \mathbf{f}_b(\mathbf{x},t),
\quad\quad
\text{on } \partial\Omega_o,
\label{equ:bc_obc_C_1}
\end{equation}
\begin{equation}
\text{\underline{OBC-E}:} \
-p\mathbf{n} + \nu\mathbf{n}\cdot\nabla\mathbf{u}
  - \left(\frac{1}{2}\left|\mathbf{u} \right|^2 \mathbf{n}  \right)     
    \Theta_0(\mathbf{n},\mathbf{u})
= \mathbf{f}_b(\mathbf{x},t),
\quad\quad
\text{on } \partial\Omega_o,
\label{equ:bc_obc_E}
\end{equation}
\begin{equation}
\text{\underline{OBC-F}:} \
-p\mathbf{n} + \nu\mathbf{n}\cdot\nabla\mathbf{u}
  - \left[\frac{1}{2} \left(\mathbf{n} \cdot \mathbf{u} \right)\mathbf{u}
    \right] \Theta_0(\mathbf{n},\mathbf{u})
= \mathbf{f}_b(\mathbf{x},t),
\quad\quad
\text{on } \partial\Omega_o,
\label{equ:bc_obc_F}
\end{equation}
where the boundary condition \eqref{equ:bc_obc_E}
is developed in \cite{DongKC2013}.
The condition \eqref{equ:bc_obc_C_1} is 
a modified form for that 
of \cite{BazilevsGHMZ2009,GravemeierCYIW2012,IsmailGCW2014}.
The condition \eqref{equ:bc_obc_F} is 
a modified form based on that of \cite{BruneauF1994}.


These conditions belong to the 
the following generalized form of open boundary condition
\begin{equation}
-p\mathbf{n} + \nu\mathbf{n}\cdot\nabla\mathbf{u}
  - \left[
      \left(\theta + \alpha_2 \right)\frac{1}{2}\left|\mathbf{u} \right|^2 \mathbf{n}  
      + \left(1-\theta + \alpha_1\right) \frac{1}{2}\left(\mathbf{n} \cdot \mathbf{u} \right)\mathbf{u}
    \right] \Theta_0(\mathbf{n},\mathbf{u})
= \mathbf{f}_b(\mathbf{x},t),
\quad\quad
\text{on } \partial\Omega_o,
\label{equ:bc_obc_gobc_1}
\end{equation}
where $\theta$, $\alpha_1$ and $\alpha_2$ are chosen constants satisfying 
the conditions
\begin{equation}
0\leqslant \theta \leqslant 1, \qquad
\alpha_1 \geqslant 0, \qquad
\alpha_2 \geqslant 0.
\end{equation}
For example, 
OBC-A corresponds to
\eqref{equ:bc_obc_gobc_1} with 
$(\theta,\alpha_1,\alpha_2)=\left(\frac{1}{2},0,0\right)$,
OBC-B 
corresponds to
\eqref{equ:bc_obc_gobc_1} with 
$(\theta,\alpha_1,\alpha_2)=\left(1,0,1\right)$, and
OBC-C corresponds to \eqref{equ:bc_obc_gobc_1} with
$(\theta,\alpha_1,\alpha_2)=\left(1,1,0\right)$.
OBC-D, OBC-E and OBC-F respectively correspond to
\eqref{equ:bc_obc_gobc_1} with
$(\theta,\alpha_1,\alpha_2)=\left(0,1,0\right)$,
$(\theta,\alpha_1,\alpha_2)=\left(1,0,0\right)$
and $(\theta,\alpha_1,\alpha_2)=\left(0,0,0\right)$.

The boundary condition \eqref{equ:bc_obc_gobc_1} is
in turn a special case of the following more general form
of open boundary condition
\begin{multline}
-p\mathbf{n} + \nu\mathbf{n}\cdot\nabla\mathbf{u}
  - \left[
      \theta \frac{1}{2}\left|\mathbf{u} \right|^2 \mathbf{n}  
      + \left(1-\theta\right) \frac{1}{2} \left(\mathbf{n} \cdot \mathbf{u} \right)\mathbf{u}
      \right. \\
      \left.
      + \beta(\mathbf{n},\mathbf{u}) \mathbf{n} \times \mathbf{u}
      - C_1(\mathbf{n},\mathbf{u}) \mathbf{u}
      + C_2(\mathbf{n},\mathbf{u}) \mathbf{n}
    \right] \Theta_0(\mathbf{n},\mathbf{u})
= \mathbf{f}_b(\mathbf{x},t),
\quad\quad
\text{on } \partial\Omega_o,
\label{equ:bc_obc_gobc_2}
\end{multline}
where $\beta(\mathbf{n},\mathbf{u})$ is an arbitrary scalar function or constant,
while $C_1(\mathbf{n},\mathbf{u})\geqslant 0$ and 
$C_2(\mathbf{n},\mathbf{u})\geqslant 0$ are two non-negative scalar functions
or constants.
The boundary condition \eqref{equ:bc_obc_gobc_1} is obtained
from \eqref{equ:bc_obc_gobc_2} by letting
$\beta(\mathbf{n},\mathbf{u}) = 0$, 
$C_1(\mathbf{n},\mathbf{u}) = - \frac{\alpha_1}{2}(\mathbf{n}\cdot\mathbf{u})$
($\alpha_1\geqslant 0$),
and $C_2(\mathbf{n},\mathbf{u}) = \frac{\alpha_2}{2}\left|\mathbf{u} \right|^2$
($\alpha_2\geqslant 0$).


To understand the rationale underlying these boundary conditions,
we consider the energy-balance equation 
for the 
system \eqref{equ:nse}--\eqref{equ:continuity},
\begin{equation}
\begin{split}
\frac{\partial}{\partial t} \int_{\Omega} \frac{1}{2}\left|\mathbf{u} \right|^2
=
& -\nu \int_{\Omega}\| \nabla\mathbf{u} \|^2 + \int_{\Omega} \mathbf{f}\cdot \mathbf{u}
  + \int_{\partial\Omega_d} \left(
              \mathbf{n}\cdot\mathbf{T}\cdot \mathbf{u}
              - \frac{1}{2}\left|\mathbf{u} \right|^2 \mathbf{n}\cdot \mathbf{u}
         \right)   \\
&
  + \int_{\partial\Omega_o} \left(
              \mathbf{n}\cdot\mathbf{T} \cdot \mathbf{u}
	      - \frac{1}{2}\left|\mathbf{u} \right|^2\mathbf{n}\cdot \mathbf{u}
         \right),
\end{split}
\label{equ:energy}
\end{equation}
where $\mathbf{T} = -p\mathbf{I} + \nu\nabla\mathbf{u}$ 
($\mathbf{I}$ is the identity tensor).
We assume $\mathbf{f}_b=0$ in
\eqref{equ:bc_obc_gobc_2} and that $\delta \rightarrow 0$ 
in the smoothed step function $\Theta_0(\mathbf{n},\mathbf{u})$. 
Employing boundary condition \eqref{equ:bc_obc_gobc_2},
one can then get
\begin{equation}
\begin{split}
\mathbf{n}\cdot\mathbf{T} \cdot \mathbf{u}
	      - \frac{1}{2}\left|\mathbf{u} \right|^2\mathbf{n}\cdot \mathbf{u}
& 
= \left(-p\mathbf{n} + \nu\mathbf{n}\cdot\nabla\mathbf{u} \right) \cdot \mathbf{u}
   - \frac{1}{2}\left|\mathbf{u} \right|^2\mathbf{n}\cdot \mathbf{u} \\
&
= \left\{
\begin{array}{ll}
- C_1\left|\mathbf{u} \right|^2
+ C_2\left(\mathbf{n}\cdot\mathbf{u} \right),
 & \text{if} \ \mathbf{n}\cdot\mathbf{u} < 0, \\
- \frac{1}{2}\left|\mathbf{u} \right|^2\left(\mathbf{n}\cdot \mathbf{u}\right), 
& \text{if} \ \mathbf{n}\cdot\mathbf{u} \geqslant 0,
\end{array}
\right.
\qquad \text{on} \ \partial\Omega_o,
\quad \text{as} \ \delta \rightarrow 0.
\end{split}
\end{equation}
Therefore, with the boundary condition \eqref{equ:bc_obc_gobc_2},
the last surface integral over the open boundary $\partial\Omega_o$
in the energy balance equation
\eqref{equ:energy} will always be non-positive
if $\delta$ is sufficiently small. 
This ensures the energy stability of the system (in the absence
of external forces),
even if there exists backflow or energy 
influx (i.e. $\mathbf{n}\cdot\mathbf{u}<0$) into
the domain through the open boundary $\partial\Omega_o$.

Apart from the boundary conditions, we assume
the following initial condition for
the velocity
\begin{equation}
\mathbf{u}(\mathbf{x},t=0) = \mathbf{u}_{in}(\mathbf{x}),
\label{equ:ic}
\end{equation}
where $\mathbf{u}_{in}$ is the initial velocity field
satisfying equation \eqref{equ:continuity} 
and compatible with the boundary
condition \eqref{equ:bc_dbc}.

The governing equations \eqref{equ:nse} and \eqref{equ:continuity},
supplemented by the boundary 
condition \eqref{equ:bc_dbc} on $\partial\Omega_d$
and an open boundary condition (among equations 
\eqref{equ:bc_obc_A}--\eqref{equ:bc_obc_D}, 
or \eqref{equ:bc_obc_C_1}--\eqref{equ:bc_obc_F}, or 
\eqref{equ:bc_obc_gobc_1})
on $\partial\Omega_o$, together with the
initial condition \eqref{equ:ic} for the velocity,
constitute the system to be solved in numerical simulations.



\subsection{Algorithm Formulation}
\label{sec:alg_pcorr_obc}


In this section 
we present an algorithm based on a pressure correction-type
strategy for solving the governing equations together
with the boundary conditions discussed above.
Our emphasis here is on the numerical treatment of the 
open boundary conditions. 

To facilitate  subsequent discussions, we re-write 
the open boundary condition \eqref{equ:bc_obc_gobc_1}
in a more compact form as follows:
\begin{equation}
-p\mathbf{n} + \nu\mathbf{n}\cdot\nabla\mathbf{u}
 - \mathbf{E}(\mathbf{n},\mathbf{u}) = \mathbf{f}_b(\mathbf{x},t),
\qquad \text{on} \ \partial\Omega_o,
\label{equ:bc_obc_gobc_1_reform}
\end{equation}
where 
\begin{equation}
\mathbf{E}(\mathbf{n},\mathbf{u}) = 
\left[
      \left(\theta + \alpha_2 \right)\frac{1}{2}\left|\mathbf{u} \right|^2 \mathbf{n}  
      + \left(1-\theta + \alpha_1\right) \frac{1}{2}\left(\mathbf{n} \cdot \mathbf{u} \right)\mathbf{u}
    \right] \Theta_0(\mathbf{n},\mathbf{u}).
\label{equ:E_expr}
\end{equation}
The system to solve consists of equations
\eqref{equ:nse} and \eqref{equ:continuity},
together with the boundary conditions
\eqref{equ:bc_dbc} and \eqref{equ:bc_obc_gobc_1_reform}.

Let $n$ ($n\geqslant 0$) denote the time step index, and 
$(\cdot)^n$ denote the variable $(\cdot)$
at time step $n$. 
We use $\tilde{\mathbf{u}}^n$ and $\mathbf{u}^n$
to denote two slightly different approximations of
the velocity $\mathbf{u}$ at step $n$.
Define
\begin{equation}
\tilde{\mathbf{u}}^0 = \mathbf{u}_{in}, \quad
\mathbf{u}^0 = \mathbf{u}_{in}.
\label{equ:ic_1}
\end{equation}

Let 
$
H_{p0}^1(\Omega) = \left\{ \
v\in H^1(\Omega) \ : \ v|_{\partial\Omega_o} = 0
\ \right\}.
$
By enforcing equation \eqref{equ:nse} at $t=0$ and using
equations \eqref{equ:continuity} and \eqref{equ:bc_dbc},
we obtain an equation in weak form about the initial
pressure $p^0$,
\begin{multline}
\int_{\Omega}\nabla p^0\cdot\nabla q =
\int_{\Omega}\left( 
\mathbf{f}^0 - \mathbf{u}_{in}\cdot\nabla\mathbf{u}_{in}
\right) \cdot \nabla q
- \nu \int_{\partial\Omega_d\cup\partial\Omega_o} \mathbf{n} \times (\nabla\times\mathbf{u}_{in})
\cdot \nabla q \\
-\int_{\partial\Omega_d}\mathbf{n}\cdot\left.\frac{\partial \mathbf{w}}{\partial t}\right|^{0} q,
\quad \forall q\in H^1_{p0}(\Omega),
\label{equ:init_p}
\end{multline}
where $q$ is a test function and
$\mathbf{n}$ is the outward-pointing unit vector
normal to $\partial\Omega$.
$\left.\frac{\partial\mathbf{w}}{\partial t} \right|^0$
denotes $\frac{\partial\mathbf{w}}{\partial t}$
at time step zero, and can be
approximated discretely (e.g. by the second-order
backward differentiation formula) because
the boundary velocity $\mathbf{w}(\mathbf{x},t)$
is known on $\partial\Omega_d$.
This equation can be solved for $p^0$,
together with the following pressure
Dirichlet condition
\begin{equation}
p^0 = \nu\mathbf{n}\cdot\nabla\mathbf{u}_{in}\cdot\mathbf{n}
- \mathbf{n}\cdot\mathbf{E}(\mathbf{n},\mathbf{u}_{in})
- \mathbf{f}_b^0\cdot\mathbf{n},
\quad \text{on} \ \partial\Omega_o.
\end{equation}

Given ($\tilde{\mathbf{u}}^n$, $\mathbf{u}^n$, $p^n$), 
we compute ($\tilde{\mathbf{u}}^{n+1}$,  $p^{n+1}$, $\mathbf{u}^{n+1}$),
together with an auxiliary scalar field variable
$\phi^{n+1}$, 
successively in a de-coupled fashion
as follows: \\ [0.1in] 
\noindent\underline{for $\tilde{\mathbf{u}}^{n+1}$:}
\begin{subequations}
\begin{equation}
\frac{\gamma_0\tilde{\mathbf{u}}^{n+1} - \hat{\mathbf{u}}}{\Delta t}
+ \tilde{\mathbf{u}}^{*,n+1}\cdot\nabla\tilde{\mathbf{u}}^{*,n+1}
+ \nabla p^n - \nu\nabla^2\tilde{\mathbf{u}}^{n+1}
= \mathbf{f}^{n+1},
\label{equ:velocity_1}
\end{equation}
\begin{equation}
\tilde{\mathbf{u}}^{n+1} = \mathbf{w}^{n+1},
\qquad \text{on} \ \partial\Omega_d,
\label{equ:velocity_2}
\end{equation}
\begin{equation}
\mathbf{n}\cdot\nabla\tilde{\mathbf{u}}^{n+1}
= \frac{1}{\nu}\left[
  p^{*,n+1}\mathbf{n} + \mathbf{E}(\mathbf{n}, \tilde{\mathbf{u}}^{*,n+1})
  + \mathbf{f}_b^{n+1}
\right],
\qquad \text{on} \ \partial\Omega_o.
\label{equ:velocity_3}
\end{equation}
\end{subequations}
\noindent\underline{for $\phi^{n+1}$:}
\begin{subequations}
\begin{equation}
\frac{\gamma_0}{\Delta t}\phi^{n+1} - \nu\nabla^2\phi^{n+1}
= \nabla\cdot\left[
  \mathbf{f}^{n+1} - \tilde{\mathbf{u}}^{*,n+1}\cdot\nabla\tilde{\mathbf{u}}^{*,n+1}
  - \nabla p^n
\right],
\label{equ:phi_1}
\end{equation}
\begin{multline}
\mathbf{n}\cdot\nabla\phi^{n+1}
= \frac{1}{\nu}\mathbf{n}\cdot\frac{\gamma_0\mathbf{w}^{n+1}-\hat{\mathbf{w}}}{\Delta t}
- \frac{1}{\nu} \mathbf{n} \cdot \left[
  \mathbf{f}^{n+1} - \tilde{\mathbf{u}}^{*,n+1}\cdot\nabla\tilde{\mathbf{u}}^{*,n+1}
  - \nabla p^n
\right] \\
+ \mathbf{n}\cdot \nabla\times\tilde{\bm{\omega}}^{n+1},
\quad \text{on} \ \partial\Omega_d,
\label{equ:phi_2}
\end{multline}
\begin{equation}
\phi^{n+1} = \nabla\cdot\tilde{\mathbf{u}}^{n+1},
\qquad \text{on} \ \partial\Omega_o.
\label{equ:phi_3}
\end{equation}
\end{subequations}
\noindent\underline{for $p^{n+1}$:}
\begin{subequations}
\begin{equation}
\frac{\gamma_0\mathbf{u}^{n+1}-\gamma_0\tilde{\mathbf{u}}^{n+1}}{\Delta t}
+ \nabla\left(p^{n+1} - p^n + \nu\phi^{n+1} \right) = 0,
\qquad \label{equ:pressure_1}
\end{equation}
\begin{equation}
\nabla\cdot\mathbf{u}^{n+1} = 0,
\label{equ:pressure_2}
\end{equation}
\begin{equation}
\mathbf{n}\cdot\mathbf{u}^{n+1} = \mathbf{n}\cdot\mathbf{w}^{n+1},
\qquad \text{on} \ \partial\Omega_d,
\label{equ:pressure_3}
\end{equation}
\begin{equation}
p^{n+1} = \nu\mathbf{n}\cdot\nabla\tilde{\mathbf{u}}^{n+1}\cdot\mathbf{n}
 - \mathbf{n}\cdot\mathbf{E}\left(\mathbf{n},\tilde{\mathbf{u}}^{n+1} \right)
 - \mathbf{f}_b^{n+1}\cdot\mathbf{n} - \nu\phi^{n+1},
\quad \text{on} \ \partial\Omega_o.
\label{equ:pressure_4}
\end{equation}
\end{subequations}
\noindent\underline{for $\mathbf{u}^{n+1}$:}
\begin{equation}
\mathbf{u}^{n+1} = \tilde{\mathbf{u}}^{n+1}
 - \frac{\Delta t}{\gamma_0} \nabla\left(
  p^{n+1} - p^n + \nu\phi^{n+1}
 \right).
\label{equ:vel_1}
\end{equation}



The meanings of the symbols involved in the above equations
\eqref{equ:velocity_1}--\eqref{equ:vel_1} are
as follows.
$\Delta t$ denotes the time step size.
Let $J$ ($J=1$ or $2$) denote the temporal order of
the scheme. Then $\tilde{\mathbf{u}}^{*,n+1}$
and $p^{*,n+1}$ respectively denote the  $J$-th
order explicit approximations of
$\tilde{\mathbf{u}}^{n+1}$ and $p^{n+1}$,
given by
\begin{equation}
\tilde{\mathbf{u}}^{*,n+1} = \left\{
\begin{array}{ll}
\tilde{\mathbf{u}}^n, & J=1, \\
2\tilde{\mathbf{u}}^n - \tilde{\mathbf{u}}^{n-1}, & J=2,
\end{array}
\right.
\qquad
p^{*,n+1} = \left\{
\begin{array}{ll}
p^n, & J=1, \\
2p^n - p^{n-1}, & J=2.
\end{array}
\right.
\end{equation} 
$\hat{\mathbf{u}}$ and the constant $\gamma_0$ are
given by
\begin{equation}
\hat{\mathbf{u}} = \left\{
\begin{array}{ll}
\mathbf{u}^n, & J=1, \\
2\mathbf{u}^n - \frac{1}{2}\mathbf{u}^{n-1}, & J=2,
\end{array}
\right.
\qquad
\gamma_0 = \left\{
\begin{array}{ll}
1, & J=1, \\
\frac{3}{2}, & J=2.
\end{array}
\right.
\end{equation}
$\mathbf{w}$ is the boundary velocity on $\partial\Omega_d$,
and $\hat{\mathbf{w}}$ is defined in the same way
as $\hat{\mathbf{u}}$ defined above.
The auxiliary variable $\phi^{n+1}$
represents an approximation of the quantity
$\nabla\cdot\tilde{\mathbf{u}}^{n+1}$.
$\mathbf{n}$ is the outward-pointing unit vector normal to
the boundary.
$\tilde{\bm{\omega}}^{n+1}$ denotes the vorticity,
$\tilde{\bm{\omega}}^{n+1} = \nabla\times\tilde{\mathbf{u}}^{n+1}$.
$\mathbf{E}(\mathbf{n},\mathbf{u})$ is defined 
in equation \eqref{equ:E_expr}.




One can recognize that the overall structure of the above algorithm 
resembles a rotational incremental pressure
correction-type strategy (see \cite{GuermondMS2006}).
Two features distinguish the above scheme from
the usual pressure correction formulations. 
One feature  lies in the introduction of the
equation \eqref{equ:phi_1} for the 
variable $\phi^{n+1}$ and the associated boundary conditions
\eqref{equ:phi_2} and \eqref{equ:phi_3}.
One can note that this equation for $\phi^{n+1}$ exists only in the 
discrete sense, and it differs from the dynamic equation
about $\nabla\cdot\mathbf{u}$ at the continuum level.
Another aspect that this scheme 
differs from the usual  
formulation lies in the form of the second term
on the left hand side (LHS) of equation \eqref{equ:pressure_1}.
This form allows us to compute the pressure $p^{n+1}$ directly 
in the $H^1(\Omega)$ space. On the other hand,
one notes that with the usual rotational pressure-correction 
formulation \cite{TimmermansMV1996,GuermondMS2006}
the pressure $p^{n+1}$ resides in the $L^2(\Omega)$ space. 
More importantly, this form allows
for a straightforward discrete pressure condition (see \eqref{equ:pressure_4})
on the open domain boundary. 
We would like to
point out that the purpose of equation \eqref{equ:vel_1}
is for the evaluation of 
$\mathbf{u}^{n+1}$ in the $L^2(\Omega)$ space,
not for the projection to the $H^1(\Omega)$ space.

We briefly mention some variants to the treatment 
of the governing equations. 
%
An alternative to the $\phi^{n+1}$ step
(equations \eqref{equ:phi_1}--\eqref{equ:phi_3})
of the above algorithm is the following,
\begin{equation}
\phi^{n+1} = \nabla\cdot\tilde{\mathbf{u}}^{n+1}.
\label{equ:project_div}
\end{equation}
This amounts to a projection of $\nabla\cdot\tilde{\mathbf{u}}$
to the $H^1(\Omega)$ space,
and requires the solution of a linear algebraic
system involving the global mass matrix.
The computational costs for solving
\eqref{equ:project_div} and for solving
\eqref{equ:phi_1}--\eqref{equ:phi_3}
are comparable. However,
we observe that the algorithm using
equations \eqref{equ:phi_1}--\eqref{equ:phi_3}
provides consistently improved accuracy 
for the pressure than that using \eqref{equ:project_div}.
%
In equations \eqref{equ:phi_1}--\eqref{equ:phi_2},
replacing 
$\tilde{\mathbf{u}}^{*,n+1}\cdot\nabla\tilde{\mathbf{u}}^{*,n+1}$
by $\tilde{\mathbf{u}}^{n+1}\cdot\nabla\tilde{\mathbf{u}}^{n+1}$
makes little difference 
in terms of stability and accuracy in numerical simulations. 
However, it increases the computational cost to a certain extent
because of the need for the extra computation
of $\tilde{\mathbf{u}}^{n+1}\cdot\nabla\tilde{\mathbf{u}}^{n+1}$.

Let us now comment on the numerical treatments of
the boundary conditions. 
In the velocity substep for $\tilde{\mathbf{u}}^{n+1}$,
we have imposed a velocity Neumann-type condition \eqref{equ:velocity_3}
on $\partial\Omega_o$, which is derived
from the open boundary condition \eqref{equ:bc_obc_gobc_1_reform}.
The pressure and velocity are treated explicitly in
this Neumann condition.
A variant form
for the velocity Neumann condition \eqref{equ:velocity_3} is
\begin{equation}
\mathbf{n}\cdot\nabla\tilde{\mathbf{u}}^{n+1}
= \frac{1}{\nu}\left[
  p^{n}\mathbf{n} + \mathbf{E}(\mathbf{n}, \tilde{\mathbf{u}}^{*,n+1})
  + \mathbf{f}_b^{n+1}
\right],
\qquad \text{on} \ \partial\Omega_o,
\label{equ:velocity_3_variant}
\end{equation}
which is also observed to be stable.
When solving for $\phi^{n+1}$, 
we have imposed a Neumann-type condition 
\eqref{equ:phi_2} on $\partial\Omega_d$
and Dirichlet-type condition \eqref{equ:phi_3} on
$\partial\Omega_o$.
In the pressure substep for $p^{n+1}$,
a pressure Dirichlet-type condition \eqref{equ:pressure_4}
has been imposed on the open boundary $\partial\Omega_o$.
This pressure condition is essentially obtained
from the open boundary condition \eqref{equ:bc_obc_gobc_1_reform},
by taking the inner product 
between this equation and $\mathbf{n}$, 
and it contains an extra term $-\nu\phi^{n+1}$.
The velocity in the pressure Dirichlet condition
is approximated using $\tilde{\mathbf{u}}^{n+1}$
computed from a previous substep.
A variant form of the 
pressure Dirichlet condition
 \eqref{equ:pressure_4}  is the following,
\begin{equation}
p^{n+1} = \nu\mathbf{n}\cdot\nabla\tilde{\mathbf{u}}^{n+1}\cdot\mathbf{n}
 - \mathbf{n}\cdot\mathbf{E}\left(\mathbf{n},\tilde{\mathbf{u}}^{n+1} \right)
 - \mathbf{f}_b^{n+1}\cdot\mathbf{n} - \nu\nabla\cdot\tilde{\mathbf{u}}^{n+1},
\quad \text{on} \ \partial\Omega_o,
\label{equ:pressure_4_variant}
\end{equation}
which is also observed to be stable.
Note that the discrete formulations \eqref{equ:pressure_4}
and \eqref{equ:pressure_4_variant} are numerically not equivalent,
because of the need for a projection to the $H^1(\partial\Omega_o)$
space (to be discussed below) 
when imposing the Dirichlet condition \eqref{equ:phi_3}
for $\phi^{n+1}$ on $\partial\Omega_o$.

The construction $-\nu \phi^{n+1}$ in \eqref{equ:pressure_4}
(or $-\nu\nabla\cdot\tilde{\mathbf{u}}^{n+1}$ in \eqref{equ:pressure_4_variant}) 
is crucial to the 
current algorithm.
If this term is absent,
assuming $\mathbf{f}_b=0$ and that no backflow occurs
at the outflow boundary (i.e. $\mathbf{n}\cdot\mathbf{u}\geqslant 0$),
then by combining equations \eqref{equ:velocity_3} and \eqref{equ:pressure_4} 
one can show that
\begin{equation}
p^{n+1}|_{\partial\Omega_o} = p^{n} |_{\partial\Omega_o} = \cdots
= p^0 |_{\partial\Omega_o}, 
\quad
\mathbf{n}\cdot\nabla\tilde{\mathbf{u}}^{n+1}|_{\partial\Omega_o} =
\mathbf{n}\cdot\nabla\tilde{\mathbf{u}}^{n}|_{\partial\Omega_o}= \cdots
= \mathbf{n}\cdot\nabla\tilde{\mathbf{u}}^{0}|_{\partial\Omega_o},
\end{equation}
leading to a numerical locking at the open boundary $\partial\Omega_o$.

\subsection{Implementation with $C^0$ Spectral Elements}



We employ high-order spectral element methods 
\cite{SherwinK1995,KarniadakisS2005,ZhengD2011} for spatial discretizations
in the current paper.
Let us next discuss how to implement the algorithm,
\eqref{equ:velocity_1}--\eqref{equ:vel_1},
using $C^0$-continuous spectral elements. 
The formulations given below without change can also be
applied to low-order finite element methods.

The main issues are posed by the terms such as 
$\nabla\cdot\left(\tilde{\mathbf{u}}^{*,n+1}\cdot\nabla\tilde{\mathbf{u}}^{*,n+1} \right)$ in \eqref{equ:phi_1} and 
$\nabla\times\tilde{\bm{\omega}}^{n+1}$ in 
\eqref{equ:phi_2}, which cannot be readily computed
in the discrete function space with $C^0$ elements.
We will derive the weak formulations
for the algorithm, and in the process treat the
trouble terms in an appropriate fashion.


Let 
$
H_{u0}^1(\Omega) = 
\left\{ \ 
v \in H^1(\Omega) \ : \ \left. v\right|_{\partial\Omega_d} = 0
\ \right\},
$
and $\varphi \in H_{u0}^1(\Omega)$ denote the test function.
By taking the $L^2$ inner product between
 $\varphi$ and the equation \eqref{equ:velocity_1},
and integrating by part, on can obtain
the weak form for $\tilde{\mathbf{u}}^{n+1}$,
\begin{multline}
\frac{\gamma_0}{\nu\Delta t}\int_{\Omega} \varphi \tilde{\mathbf{u}}^{n+1}
+ \int_{\Omega} \nabla\varphi\cdot\nabla \tilde{\mathbf{u}}^{n+1}
= \frac{1}{\nu} \int_{\Omega}\left[
  \mathbf{f}^{n+1}
  - \tilde{\mathbf{u}}^{*,n+1}\cdot\nabla \tilde{\mathbf{u}}^{*,n+1}
  - \nabla p^n
  + \frac{\hat{\mathbf{u}}}{\Delta t}
\right] \varphi \\
+ \frac{1}{\nu}\int_{\partial\Omega_o}\left[
  p^{*,n+1}\mathbf{n}
  + \mathbf{E}(\mathbf{n},\tilde{\mathbf{u}}^{*,n+1})
  + \mathbf{f}_b^{n+1}
\right] \varphi,
\qquad
\forall \varphi \in H_{u0}^1(\Omega),
\label{equ:u_tilde_weakform}
\end{multline}
where we have used the boundary condition
\eqref{equ:velocity_3}, and
the fact that 
$
\int_{\partial\Omega_d} \mathbf{n}\cdot\nabla\tilde{\mathbf{u}}^{n+1}\varphi =0
$
because $\varphi \in H_{u0}^1(\Omega)$.

Let 
$\vartheta \in H_{p0}^1(\Omega)$ denote 
a test function.
By taking the $L^2$ inner product between $\vartheta$ and
the equation \eqref{equ:phi_1} and integrating by
part, we can get the weak form about $\phi^{n+1}$,
\begin{multline}
\frac{\gamma_0}{\nu\Delta t}\int_{\Omega} \phi^{n+1}\vartheta
+ \int_{\Omega} \nabla\phi^{n+1}\cdot\nabla\vartheta
= -\frac{1}{\nu} \int_{\Omega}\left[
  \mathbf{f}^{n+1}
  - \tilde{\mathbf{u}}^{*,n+1}\cdot\nabla \tilde{\mathbf{u}}^{*,n+1}
  - \nabla p^n
\right] \cdot \nabla\vartheta \\
+ \frac{1}{\nu} \int_{\partial\Omega_d} \mathbf{n} \cdot
     \frac{\gamma_0\mathbf{w}^{n+1}-\hat{\mathbf{w}}}{\Delta t} \vartheta
+ \int_{\partial\Omega_d}\mathbf{n}\times\tilde{\bm{\omega}}^{n+1}
       \cdot \nabla\vartheta
+ \int_{\partial\Omega_o}\mathbf{n}\times\tilde{\bm{\omega}}^{n+1}
       \cdot \nabla\vartheta,
\quad \forall \vartheta \in H_{p0}(\Omega),
\label{equ:phi_weakform}
\end{multline}
%
%
%
where we have used the divergence theorem,
the boundary condition \eqref{equ:phi_2}, and
the following identity,
\begin{equation}
\begin{split}
\int_{\partial\Omega_d} \mathbf{n}\cdot \nabla\times\tilde{\bm{\omega}}^{n+1}\vartheta
&
= \int_{\partial\Omega} \mathbf{n}\cdot \nabla\times\tilde{\bm{\omega}}^{n+1}\vartheta
= \int_{\Omega} \nabla\times\tilde{\bm{\omega}}^{n+1}\cdot\nabla\vartheta
= \int_{\Omega} \nabla\cdot\left(\tilde{\bm{\omega}}^{n+1}\times\nabla\vartheta \right) 
\\
&
= \int_{\partial\Omega} \mathbf{n} \times \tilde{\bm{\omega}}^{n+1} \cdot
       \nabla\vartheta,
\qquad \forall \vartheta \in H_{p0}(\Omega).
\end{split}
\end{equation}

Let $q\in H_{p0}^{1}(\Omega)$ denote the test function.
Taking the $L^2$ inner product between equation \eqref{equ:pressure_1}
and $\nabla q$, and integrating by part, we obtain the weak form for
$p^{n+1}$
\begin{equation}
\int_{\Omega} \nabla p^{n+1}\cdot\nabla q
= \int_{\Omega} \left[
  \frac{\gamma_0}{\Delta t} \tilde{\mathbf{u}}^{n+1}
  + \nabla\left(
      p^n - \nu\phi^{n+1}
    \right)
\right] \cdot \nabla q
- \frac{\gamma_0}{\Delta t}\int_{\partial\Omega_d} \mathbf{n}\cdot \mathbf{w}^{n+1} q,
\qquad \forall q \in H_{p0}^1(\Omega),
\label{equ:p_weakform}
\end{equation}
where we have used equations \eqref{equ:pressure_2}
and \eqref{equ:pressure_3}, and the
fact that 
$
\int_{\partial\Omega_o} \mathbf{n}\cdot\mathbf{u}^{n+1} q = 0
$
because $q\in H_{p0}^1(\Omega)$.


The weak formulations \eqref{equ:u_tilde_weakform},
\eqref{equ:phi_weakform} and \eqref{equ:p_weakform}
contain no complicating terms with derivatives
of order two or higher. All terms involved therein
can be computed directly in the discrete space
of $C^0$ elements.
These weak forms can be
discretized using $C^0$ spectral elements (or
finite elements).


Let $\Omega_h$ denote the domain
$\Omega$ partitioned using a spectral element mesh,
and $\partial\Omega_h$ denote the boundary of
$\Omega_h$,
$\partial\Omega_h = \partial\Omega_{dh}\cup\partial\Omega_{oh}$,
where $\partial\Omega_{dh}$ and $\partial\Omega_{oh}$ are
respectively the discretized $\partial\Omega_d$ and $\partial\Omega_o$.
We use $X_h\subset [H^1(\Omega_h)]^{d}$ ($d=2$ or $3$ is
the spatial dimension)
to denote the approximation space for 
the velocity $\tilde{\mathbf{u}}_h^{n+1}$,
and $M_h\subset H^1(\Omega_h)$
to denote the approximation space for the
pressure $p_h^{n+1}$ and the
field variable $\phi_h^{n+1}$.
Let 
$
X_{h0} = \{ \
v \in X_h \ : \ v|_{\partial\Omega_{dh}} = 0
\ \},
$
and
$
M_{h0} = \{ \
v \in M_h \ : \ v|_{\partial\Omega_{oh}} = 0
\ \}.
$
Then the fully discretized 
equations for \eqref{equ:u_tilde_weakform}
and \eqref{equ:velocity_2} are:
find 
$\tilde{\mathbf{u}}_h^{n+1} \in X_h$
such that
\begin{multline}
\frac{\gamma_0}{\nu\Delta t}\int_{\Omega_h} \varphi_h \tilde{\mathbf{u}}_h^{n+1}
+ \int_{\Omega_h} \nabla\varphi_h\cdot\nabla \tilde{\mathbf{u}}_h^{n+1}
= \frac{1}{\nu} \int_{\Omega_h}\left[
  \mathbf{f}_h^{n+1}
  - \tilde{\mathbf{u}}_h^{*,n+1}\cdot\nabla \tilde{\mathbf{u}}_h^{*,n+1}
  - \nabla p_h^n
  + \frac{\hat{\mathbf{u}}_h}{\Delta t}
\right] \varphi_h \\
+ \frac{1}{\nu}\int_{\partial\Omega_{oh}}\left[
  p_h^{*,n+1}\mathbf{n}_h
  + \mathbf{E}(\mathbf{n}_h,\tilde{\mathbf{u}}_h^{*,n+1})
  + \mathbf{f}_{bh}^{n+1}
\right] \varphi_h,
\qquad
\forall \varphi_h \in X_{h0},
\label{equ:u_tilde_weakform_B}
\end{multline}
and
\begin{equation}
\tilde{\mathbf{u}}_h^{n+1} = \mathbf{w}_h, \quad
\text{on} \ \partial\Omega_{dh},
\label{equ:velocity_2_B}
\end{equation}
where the subscript $(\cdot)_h$ represents the discretized version
of $(\cdot)$.
The fully discretized equations for \eqref{equ:phi_weakform}
and \eqref{equ:phi_3} are: 
find $\phi_h^{n+1} \in M_h$ such that
\begin{multline}
\frac{\gamma_0}{\nu\Delta t}\int_{\Omega_h} \phi_h^{n+1}\vartheta_h
+ \int_{\Omega_h} \nabla\phi_h^{n+1}\cdot\nabla\vartheta_h
= -\frac{1}{\nu} \int_{\Omega_h}\left[
  \mathbf{f}_h^{n+1}
  - \tilde{\mathbf{u}}_h^{*,n+1}\cdot\nabla \tilde{\mathbf{u}}_h^{*,n+1}
  - \nabla p_h^n
\right] \cdot \nabla\vartheta_h \\
+ \frac{1}{\nu} \int_{\partial\Omega_{dh}} \mathbf{n}_h \cdot
     \frac{\gamma_0\mathbf{w}_h^{n+1}-\hat{\mathbf{w}}_h}{\Delta t} \vartheta_h
+ \int_{\partial\Omega_{dh}\cup\partial\Omega_{oh}}
          \mathbf{n}_h\times\tilde{\bm{\omega}}_h^{n+1}
       \cdot \nabla\vartheta_h,
\quad \forall \vartheta_h \in M_{h0},
\label{equ:phi_weakform_B}
\end{multline}
and 
\begin{equation}
\phi_h^{n+1} = \nabla\cdot\tilde{\mathbf{u}}_h^{n+1},
\quad \text{on} \ \partial\Omega_{oh}.
\label{equ:phi_3_B}
\end{equation}
The fully discretized equations of \eqref{equ:p_weakform}
and \eqref{equ:pressure_4} are: 
find $p_h^{n+1}\in M_h$ such that
\begin{equation}
\int_{\Omega_h} \nabla p_h^{n+1}\cdot\nabla q_h
= \int_{\Omega_h} \left[
  \frac{\gamma_0}{\Delta t} \tilde{\mathbf{u}}_h^{n+1}
  + \nabla\left(
      p_h^n - \nu\phi_h^{n+1}
    \right)
\right] \cdot \nabla q_h
- \frac{\gamma_0}{\Delta t}\int_{\partial\Omega_{dh}} \mathbf{n}_h\cdot \mathbf{w}_h^{n+1} q_h,
\quad \forall q_h \in M_{h0}, 
\label{equ:p_weakform_B}
\end{equation}
and
\begin{equation}
p_h^{n+1} = \nu\mathbf{n}_h\cdot\nabla\tilde{\mathbf{u}}_h^{n+1}\cdot\mathbf{n}_h
 - \mathbf{n}_h\cdot\mathbf{E}\left(\mathbf{n}_h,\tilde{\mathbf{u}}_h^{n+1} \right)
 - \mathbf{f}_{bh}^{n+1}\cdot\mathbf{n}_h - \nu\phi_h^{n+1},
\quad \text{on} \ \partial\Omega_{oh}.
\label{equ:pressure_4_B}
\end{equation}
In addition, $\mathbf{u}_h^{n+1}$ is evaluated by
the following discretized version of equation
\eqref{equ:vel_1},
\begin{equation}
\mathbf{u}_h^{n+1} = \tilde{\mathbf{u}}_h^{n+1}
 - \frac{\Delta t}{\gamma_0} \nabla\left(
  p_h^{n+1} - p_h^n + \nu\phi_h^{n+1}
 \right).
\label{equ:vel_1_B}
\end{equation}

The final solution procedure can  therefore be summarized
as follows. Given 
($\tilde{\mathbf{u}}_h^n$, $\mathbf{u}_h^n$, $p_h^n$),
we employ the following steps to compute
the variables at time step ($n+1$):
\begin{itemize} 

\item
Solve equation \eqref{equ:u_tilde_weakform_B}, together
with the velocity Dirichlet condition \eqref{equ:velocity_2_B}
on $\partial\Omega_{dh}$,
for $\tilde{\mathbf{u}}_h^{n+1}$;

\item
Solve equation \eqref{equ:phi_weakform_B},
together with the Dirichlet condition \eqref{equ:phi_3_B}
for $\phi_h^{n+1}$
on $\partial\Omega_{oh}$,
for $\phi_h^{n+1}$;

\item
Solve equation \eqref{equ:p_weakform_B},
together with the pressure Dirichlet condition
\eqref{equ:pressure_4_B} on $\partial\Omega_{oh}$,
for $p_h^{n+1}$.

\item
Evaluate $\mathbf{u}_h^{n+1}$ based on equation
\eqref{equ:vel_1_B}, using $\tilde{\mathbf{u}}_h^{n+1}$,
$\phi_h^{n+1}$ and $p_h^{n+1}$ computed above.

\end{itemize}
It can be observed that in equation \eqref{equ:u_tilde_weakform_B}
different components of the velocity $\tilde{\mathbf{u}}_h^{n+1}$
are not coupled and therefore can be computed 
individually.


We briefly comment on how to impose the Dirichlet conditions
for $\phi_h^{n+1}$ and $p_h^{n+1}$ on the open
boundary $\partial\Omega_{oh}$ in the second and third steps 
of the above algorithm.
The expressions for the boundary conditions \eqref{equ:phi_3_B}
and \eqref{equ:pressure_4_B} both involve derivatives
of the velocity $\tilde{\mathbf{u}}_h^{n+1}$.
Consequently, the Dirichlet data for
$\phi_h^{n+1}$ and $p_h^{n+1}$ on $\partial\Omega_{oh}$
computed from \eqref{equ:phi_3_B} and \eqref{equ:pressure_4_B}
may not be continuous across the element boundaries
on $\partial\Omega_{oh}$ with $C^0$ spectral 
elements (or finite elements).
Therefore, when imposing these Dirichlet conditions,
one needs to first project the Dirichlet data computed
from \eqref{equ:phi_3_B} and \eqref{equ:pressure_4_B}
into the $H^1(\partial\Omega_{oh})$ space, and then
use the projected data for the Dirichlet conditions
on $\partial\Omega_{oh}$.
This projection essentially amounts to solving a small
linear algebraic system with the coefficient matrix
being the mass matrix on $\partial\Omega_{oh}$.
If $\partial\Omega_{oh}$ consists of several disjoint pieces,
the projection can be performed on each individual piece
separately.


As is well-known,  the approximation spaces for 
the discrete velocity and pressure should satisfy 
an inf-sup condition for compatibility, otherwise
spurious pressure modes may result.
On the other hand,
substantial evidence exists based on the works
of a number of researchers that
several types of schemes can work properly with approximation
spaces that do not satisfy the usual inf-sup condition,
e.g. with the equal-order approximation for 
the velocity and pressure;
see e.g. \cite{KarniadakisIO1991,TimmermansMV1996,
GuermondS2003,KarniadakisS2005,
GuermondMS2006,LiuLP2007,Liu2009,DongS2010,DongKC2013}
among others. 
Extensive numerical experiments of ours show that
the current splitting scheme represented by the
equations \eqref{equ:u_tilde_weakform_B}--\eqref{equ:vel_1_B}
using spectral element discretizations
can work properly with equal-order approximations for
the velocity and the pressure. No
spurious modes for the pressure are
observed.
In the current implementation and
in all the flow tests of Section \ref{sec:tests},
we have used the same orders of expansion polynomials
to approximate the velocity and the pressure
in the spectral element discretization.


There are two approximations, $\tilde{\mathbf{u}}^{n}$
and $\mathbf{u}^n$, for the velocity from the above algorithm.
The issue of which one to use in simulations has been discussed
in detail by \cite{GuermondMS2006}.
As shown by the analysis of \cite{GuermondS2004}
and pointed out by \cite{GuermondMS2006},
the two approximation velocities have the same
error estimates and in terms of
accuracy there is no reason for preferring
one to the other.
In the current paper we will use the 
approximation velocity $\tilde{\mathbf{u}}^n$
when presenting results.
All the results in Section \ref{sec:tests}
regarding the velocity are with $\tilde{\mathbf{u}}^n$.

\section{Representative Numerical Examples}
\label{sec:tests}

In this section we use several flow problems
in two dimensions (2-D)  involving inflow/outflow
boundaries to demonstrate the performance
of the numerical algorithm and the effectiveness
of the open boundary conditions developed in the
previous section. 
The flow regimes covered by the 2-D 
 simulations
range from low  to
 quite high Reynolds numbers, at which strong
backflows or vortices occur at 
the outflow/open boundaries and the physical flow
in reality would have become three-dimensional.
We compare our simulation results with
experimental data and also with the other numerical
simulations from the literature.

\subsection{Convergence Rates}
\label{sec:convergence}

The goal of this subsection is to use an
 analytic flow problem to show
the spatial and temporal convergence rates
of the method developed here.

\begin{figure}
\centerline{
\includegraphics[height=2.2in]{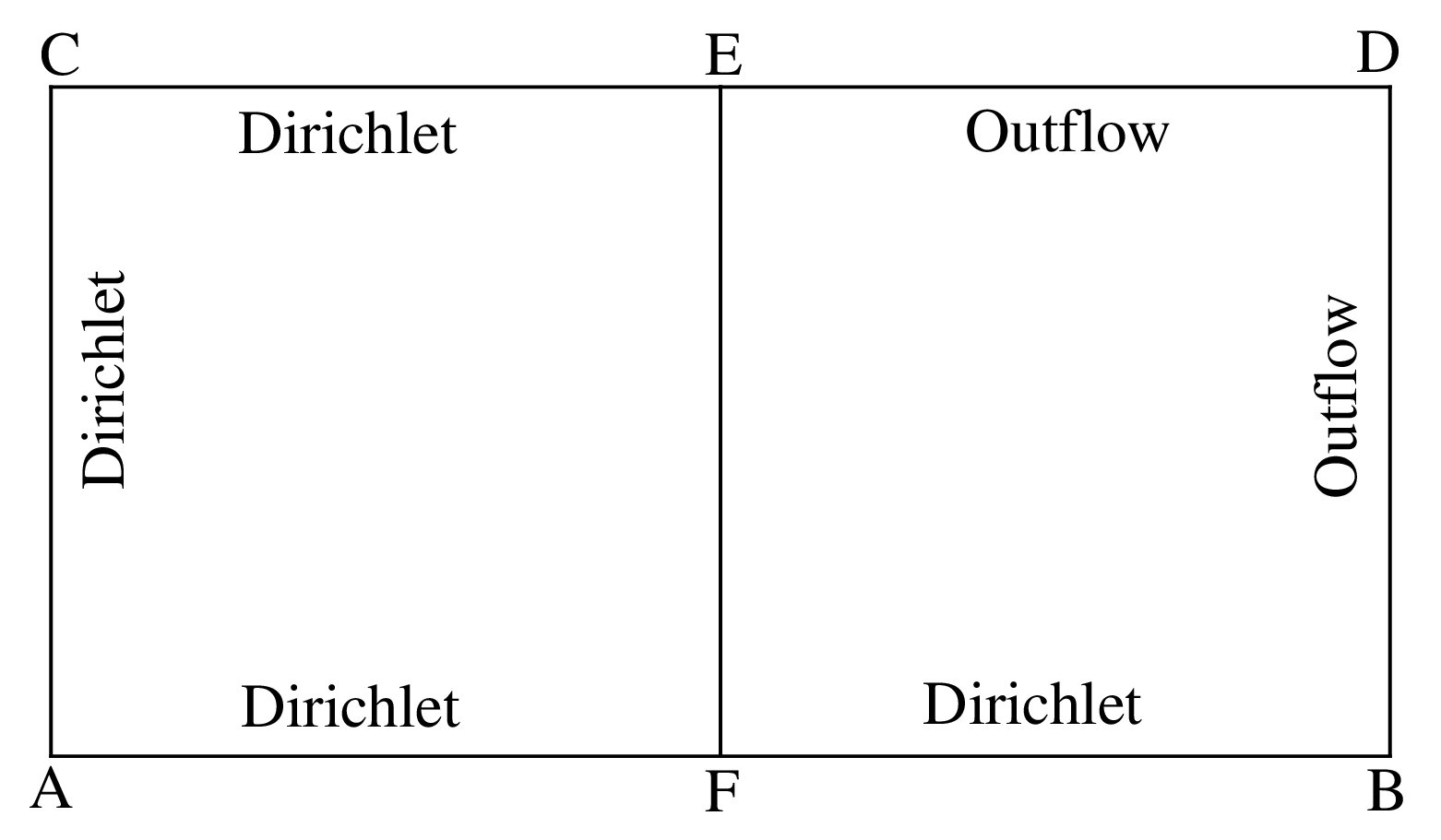}(a)
}
\centerline{
\includegraphics[height=2.8in]{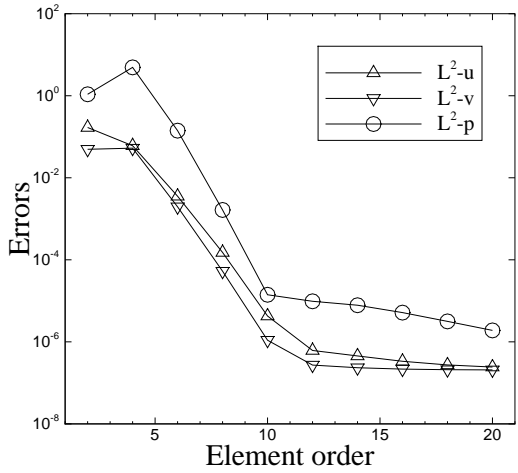}(b)
\includegraphics[height=2.8in]{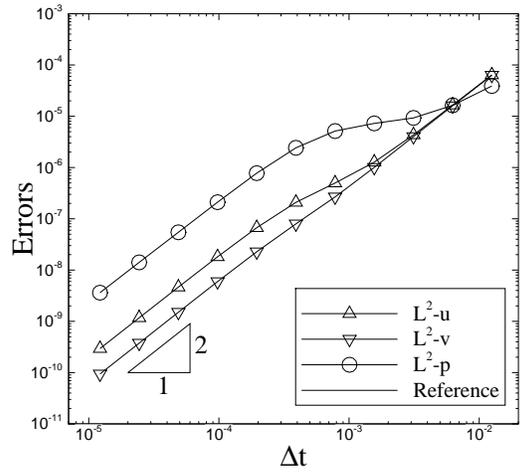}(c)
}
\caption{
Spatial and temporal convergence rates:
(a) Flow configuration and boundary conditions;
$L^2$ errors of the flow variables as a function of
the element order (b), and as a function of 
the time step size
$\Delta t$ (c). In (b) the time step size
is fixed at $\Delta t=0.001$. In (c) the element
order is fixed at $18$.
Results are obtained with 
the OBC-E outflow boundary condition.
}
\label{fig:conv}
\end{figure}

We consider the rectangular flow domain $\overline{ABCD}$
as sketched in 
Figure \ref{fig:conv}(a),
$0\leqslant x\leqslant 2$ and 
$-1 \leqslant y\leqslant 1$,
and the following analytic
expressions for the flow variables
\begin{equation}
\left\{
\begin{split}
&
u = A\cos \pi y \sin ax \sin b t, \\
&
v = -\frac{Aa}{\pi}\sin \pi y \cos ax \sin bt, \\
&
p = A \sin \pi y \sin ax \cos bt,
\end{split}
\right.
\label{equ:anal_expr}
\end{equation}
where 
$(u,v)$ are the $x$ and $y$ components of
the velocity $\mathbf{u}$, and
$A$, $a$ and $b$ are prescribed
constants whose values to be given below.
The above velocity expression satisfies
the continuity equation \eqref{equ:continuity}.
The external body force $\mathbf{f}(\mathbf{x},t)$
in \eqref{equ:nse} is chosen such that
the expressions in \eqref{equ:anal_expr}
satisfy the equation \eqref{equ:nse}.

The domain is partitioned into
two equal-sized spectral elements
$\overline{AFEC}$ and $\overline{FBDE}$
along the $x$ direction, see Figure \ref{fig:conv}(a).
On the faces $\overline{AB}$, $\overline{AC}$
and $\overline{CE}$ 
the velocity Dirichlet boundary condition \eqref{equ:bc_dbc}
is imposed, where the boundary
velocity $\mathbf{w}$ is chosen according to
the analytical expression 
from \eqref{equ:anal_expr}.
On the faces $\overline{BD}$ and $\overline{DE}$
the outflow boundary condition 
\eqref{equ:bc_obc_gobc_1_reform}
is imposed, where $\mathbf{f}_b$ is
chosen such that the analytic
expressions in \eqref{equ:anal_expr}
satisfy the equation \eqref{equ:bc_obc_gobc_1_reform}
on these boundaries.


\begin{table}
\begin{center}
\begin{tabular*}{0.6\textwidth}{ l @{\extracolsep{\fill}} c  | @{\extracolsep{\fill}} l  c  }
\hline 
parameter & value & parameter & value \\
$A$ & $2.0$ & $\delta$ & $0.05$ \\
$a$ & $\pi$ & $U_0$ & $1.0$ \\
$b$ & $1.0$ & $J$ (temporal order) & $2$  \\
$\nu$ & $0.01$ \\
\hline
\end{tabular*}
\end{center}
\caption{
Physical and numerical parameters for convergence-rate tests.
}
\label{tab:conv}
\end{table}


We employ the algorithm presented
in Section \ref{sec:alg_pcorr_obc}
to integrate the Navier-Stokes
equations in time from $t=0$
to $t=t_f$ ($t_f$ is the final time
to be given below), and
then compute the errors of the numerical
solution at $t=t_f$ against
the analytic solution given in
\eqref{equ:anal_expr}.
The element order or the time step
size $\Delta t$ is varied systematically,
and the numerical errors are monitored.
Table \ref{tab:conv} lists
the physical and numerical parameters
involved in this problem.


In the first group of tests, we fix
the time step size at $\Delta t=0.001$
and the final integration time
at $t_f=0.1$ (i.e. $100$ time steps),
and vary the element order
systematically from $2$ to $20$.
Figure \ref{fig:conv}(b) 
shows the $L^2$ errors of the 
flow variables at $t=t_f$
as a function of the element order.
These results are obtained 
with the outflow boundary condition 
OBC-E,
corresponding to the parameters
$(\theta,\alpha_1,\alpha_2)=(1,0,0)$.
One can observe that the errors
decrease exponentially as the
element order increases while below 
order $10$.
As the element order increases further beyond
$12$, the errors remain essentially
constant or decrease only slightly,
because of the saturation by
the temporal truncation errors.
These results demonstrate the spatial
exponential convergence rate
of the our method.


In the second group of tests,
we fix the final integration time at
$t_f=0.2$ and the element order
at $18$,
and vary the time step size
systematically between
$\Delta t = 1.220703125\times 10^{-5}$
and $\Delta t=0.0125$.
Figure \ref{fig:conv}(c)
shows the $L^2$ errors
of the flow variables 
as a function of $\Delta t$
 in logarithmic scales.
These results again correspond
to the outflow condition OBC-E.
On can observe
a second-order convergence rate
in time for the flow
variables as $\Delta t$
becomes small.


\subsection{Flow Past a Circular Cylinder}
\label{sec:cyl}


\begin{figure}
\centerline{
\includegraphics[width=3in]{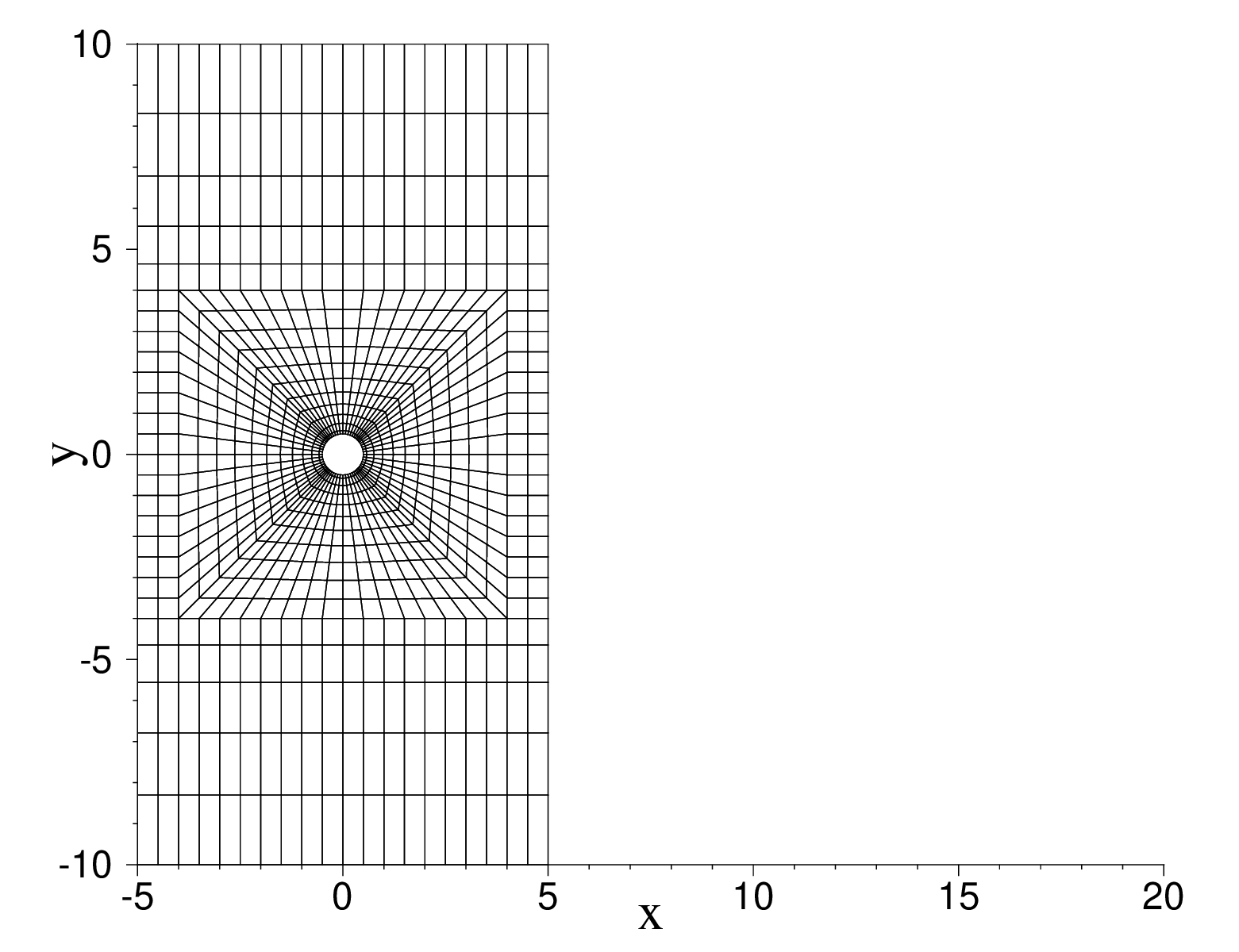}(a)
\includegraphics[width=3in]{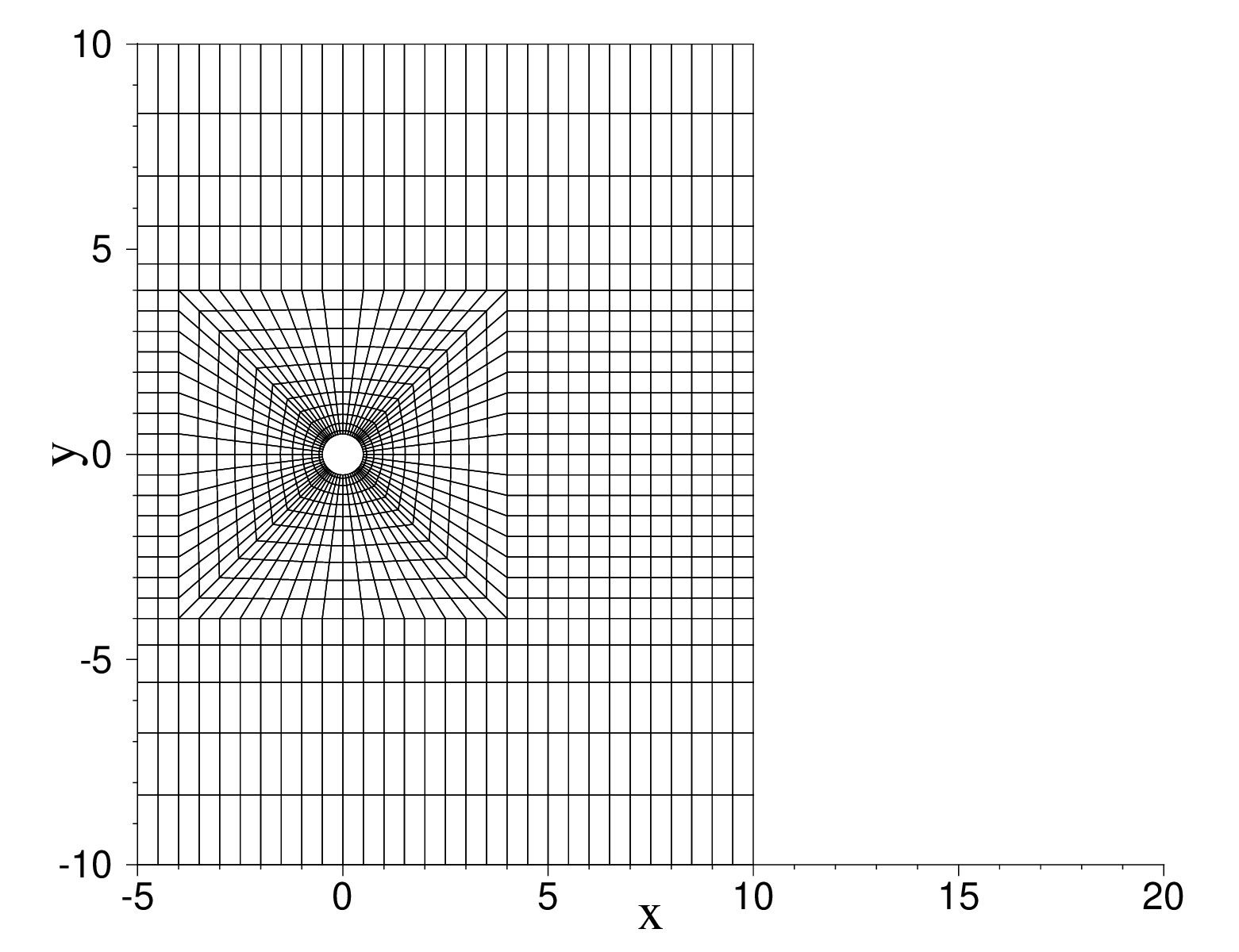}(b)
}
\centerline{
\includegraphics[width=3in]{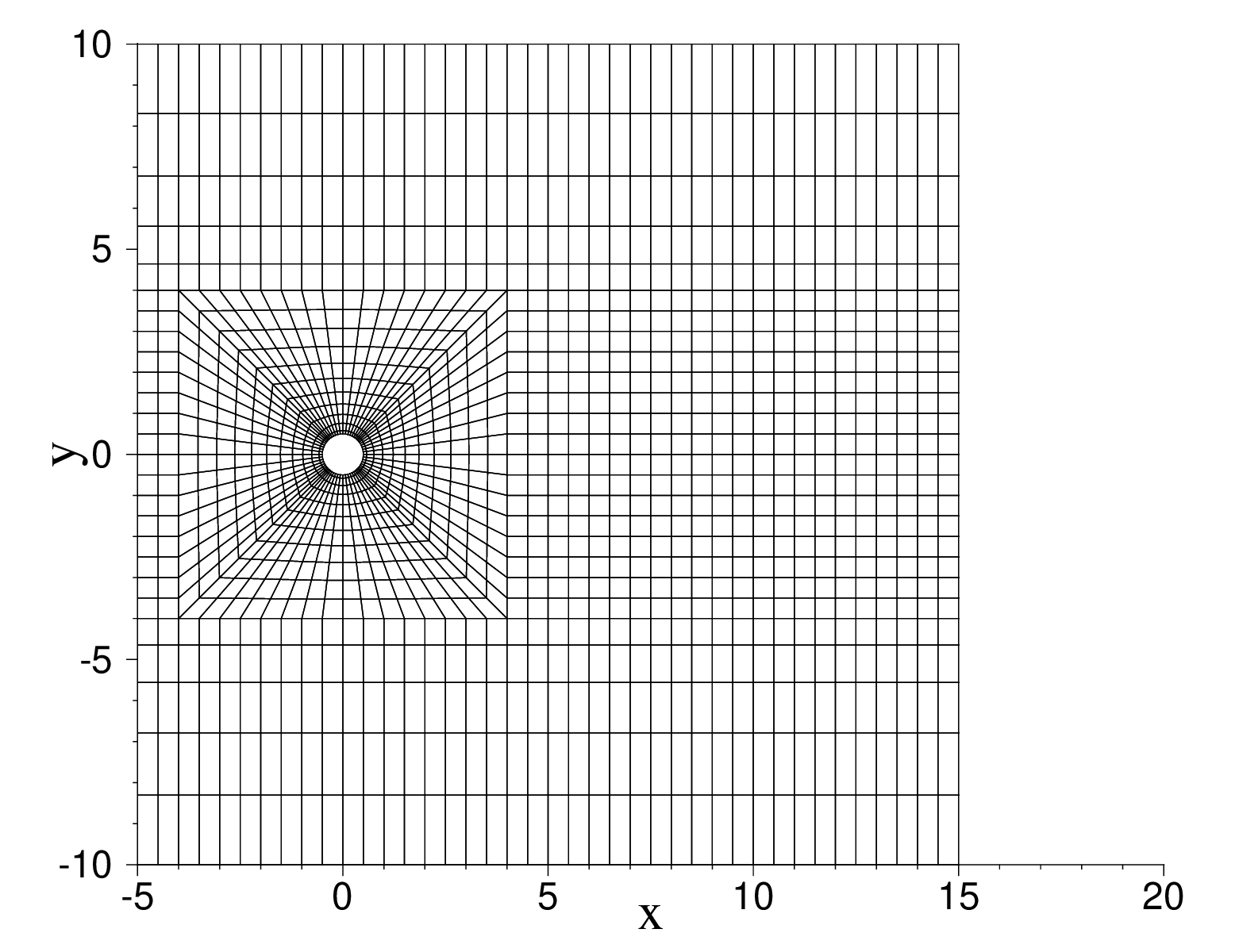}(c)
\includegraphics[width=3in]{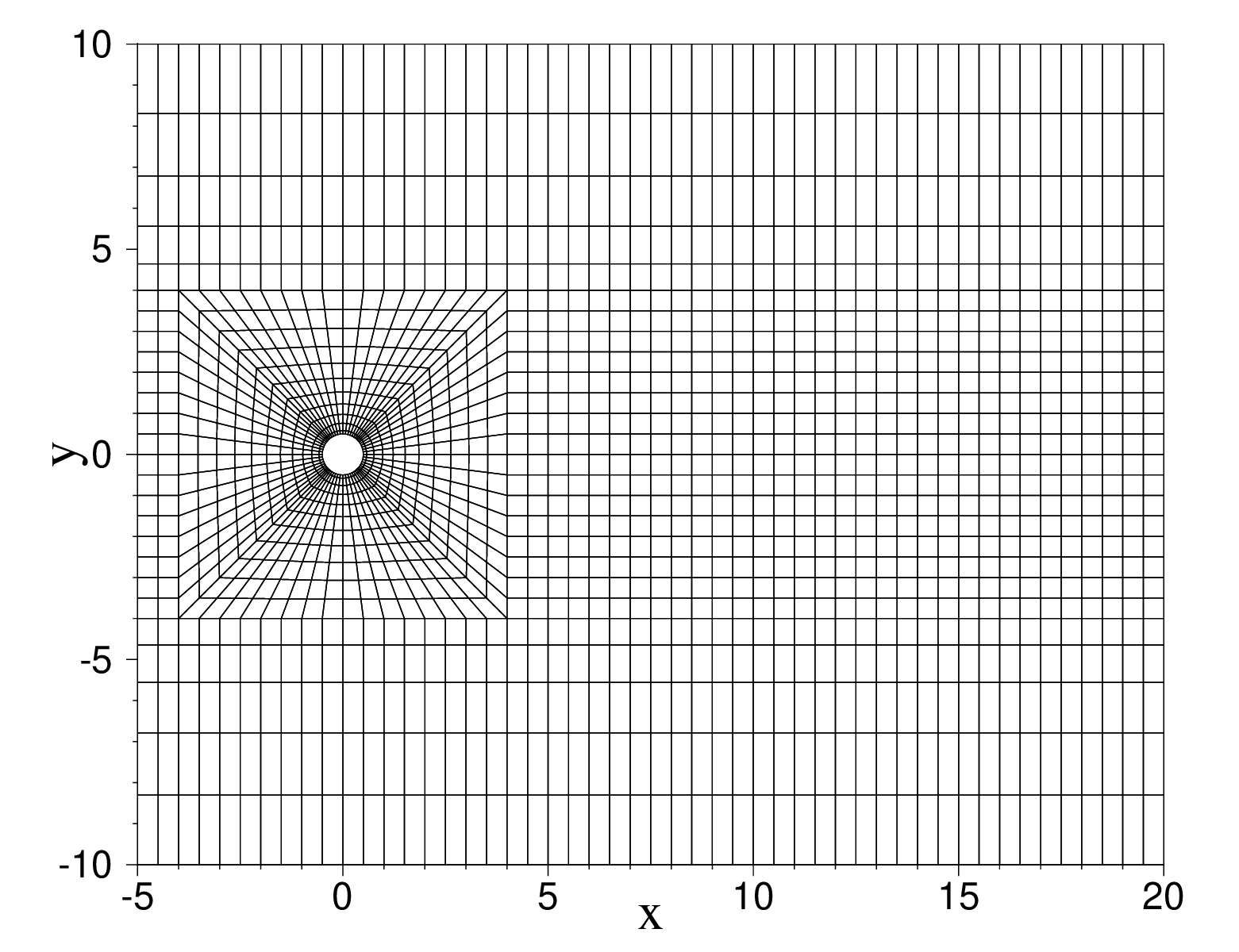}(d)
}
\caption{ Circular cylinder flow:
Flow domains of different sizes and spectral-element meshes with
(a) $968$, 
(b) $1228$,
(c) $1488$, and
(d) $1748$
quadrilateral elements. 
}
\label{fig:cyl_mesh}
\end{figure}

In this section we consider the canonical flow
past a circular cylinder in a range of Reynolds
numbers. In particular, we compare the simulation
results obtained using our method 
 with the experimental measurements
and also with other simulations from the
literature. We also demonstrate
the stability of our algorithm at high 
Reynolds numbers when strong vortices or
backflows occur at the outflow
boundaries.


The problem setting is as follows.
Consider a circular cylinder (or disk) of diameter $D$,
and the flow around the cylinder in a 
rectangular domain (see Figure \ref{fig:cyl_mesh}),
$-5D\leqslant x\leqslant L$ and 
$-10D\leqslant y\leqslant 10D$,
where $L$ is the length of the wake region to be specified
subsequently.
A uniform inflow, with a velocity
along the horizontal direction and of unit magnitude,
enters the domain through 
the left boundary ($x=-5D$).
The flow leaves the domain on the right side ($x=L$).
On the top and bottom sides of
the domain we assume that the flow is periodic.
So the configuration in practice corresponds
to the flow past an array of circular cylinders.


We have considered four domain sizes 
corresponding to
$L=5D$, $10D$, $15D$ and $20D$ (Figure \ref{fig:cyl_mesh}).
The majority of  simulations 
are performed on the domain with $L=10D$,
and simulations on the other domains have also been
conducted at several selected Reynolds numbers.
We define the Reynolds number as
\begin{equation}
Re=\frac{1}{\nu}=\frac{U_0 D}{\nu_f}
\label{equ:Re_expr}
\end{equation}
where $U_0=1$ is the free-stream inflow velocity,
 $\nu_f$ is the kinematic viscosity of
the fluid, and $\nu$ is the non-dimensional
viscosity as defined in Section \ref{sec:method}.
The Reynolds numbers covered 
in the current simulations range from
$Re=20$ to $Re=5000$. 
All the length variables are normalized by
$L$, and all the velocity variables are normalized
by $U_0$.


The flow domains have been discretized using several
spectral element meshes. 
Corresponding to the four domain sizes,
the meshes respectively consist of $968$, $1228$,
$1488$ and $1748$ quadrilateral
elements; see Figure \ref{fig:cyl_mesh}.
%
On the left domain boundary we impose
the Dirichlet condition \eqref{equ:bc_dbc},
where the boundary velocity is set to
$\mathbf{w}=(U_0,0)$.
On the top and bottom boundaries ($y/D=\pm 10$)
the periodic condition is imposed.
On the right boundary ($x=L$)
we impose the open boundary condition
\eqref{equ:bc_obc_gobc_1_reform}
with $\mathbf{f}_b=0$ and $\delta=0.01$.


We employ the algorithm developed
in Section \ref{sec:method}
for marching in time.
An element order $6$ has been used
for each element at low Reynolds numbers
(below $Re=100$), and order $8$ has been
used for each element for higher 
Reynolds numbers. We have monitored 
the forces on the cylinder. The
numerical experiments
indicate that the drag and lift coefficients
essentially do not change any more or only change slightly
when we increase the element
order further.
We use a time step size $\Delta t=10^{-3}$
for Reynolds numbers below $100$,
and $\Delta t=2.5 \times 10^{-4}$
for higher Reynolds numbers.


\begin{figure}
\centerline{
\includegraphics[width=2in]{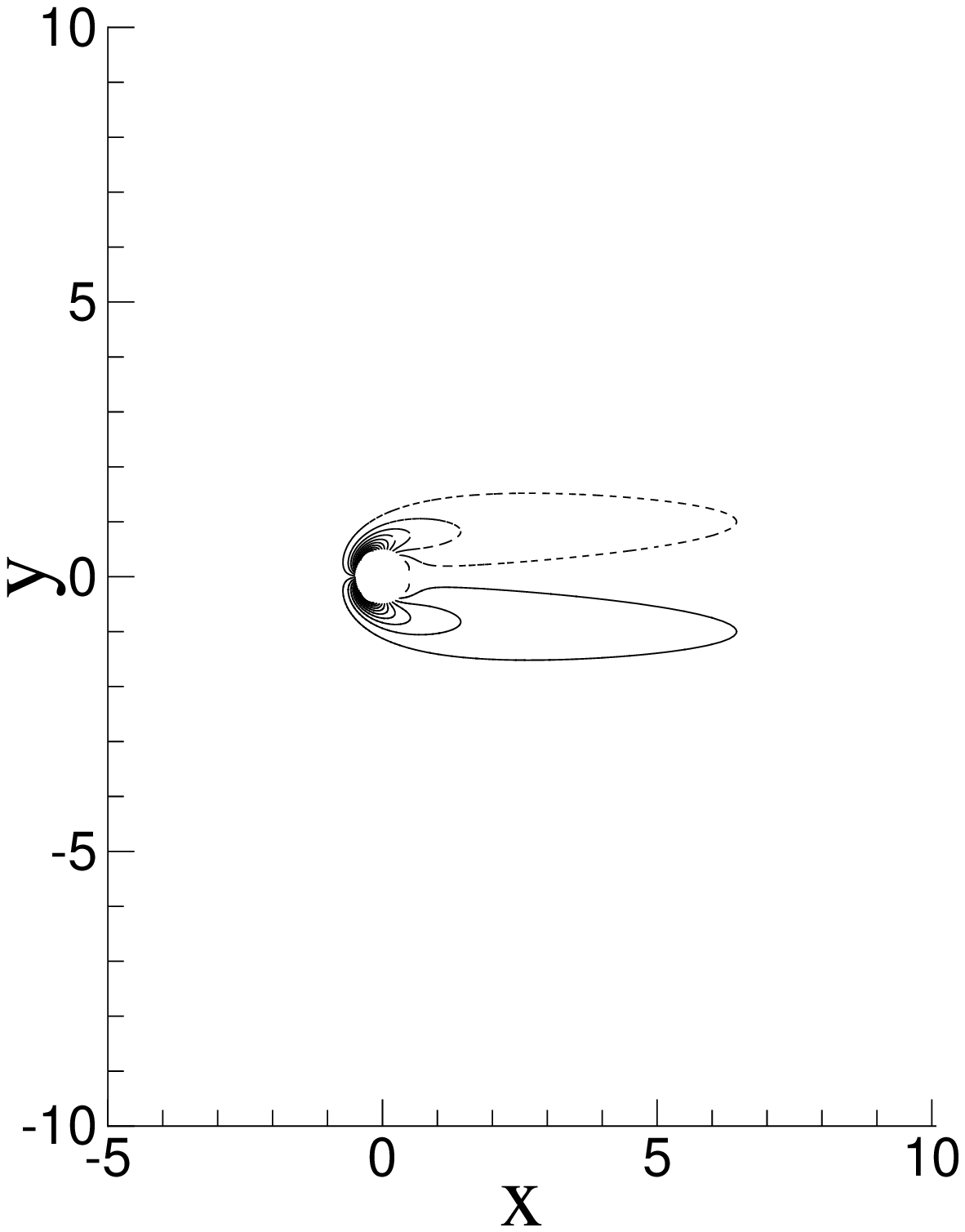}(a)
\includegraphics[width=2in]{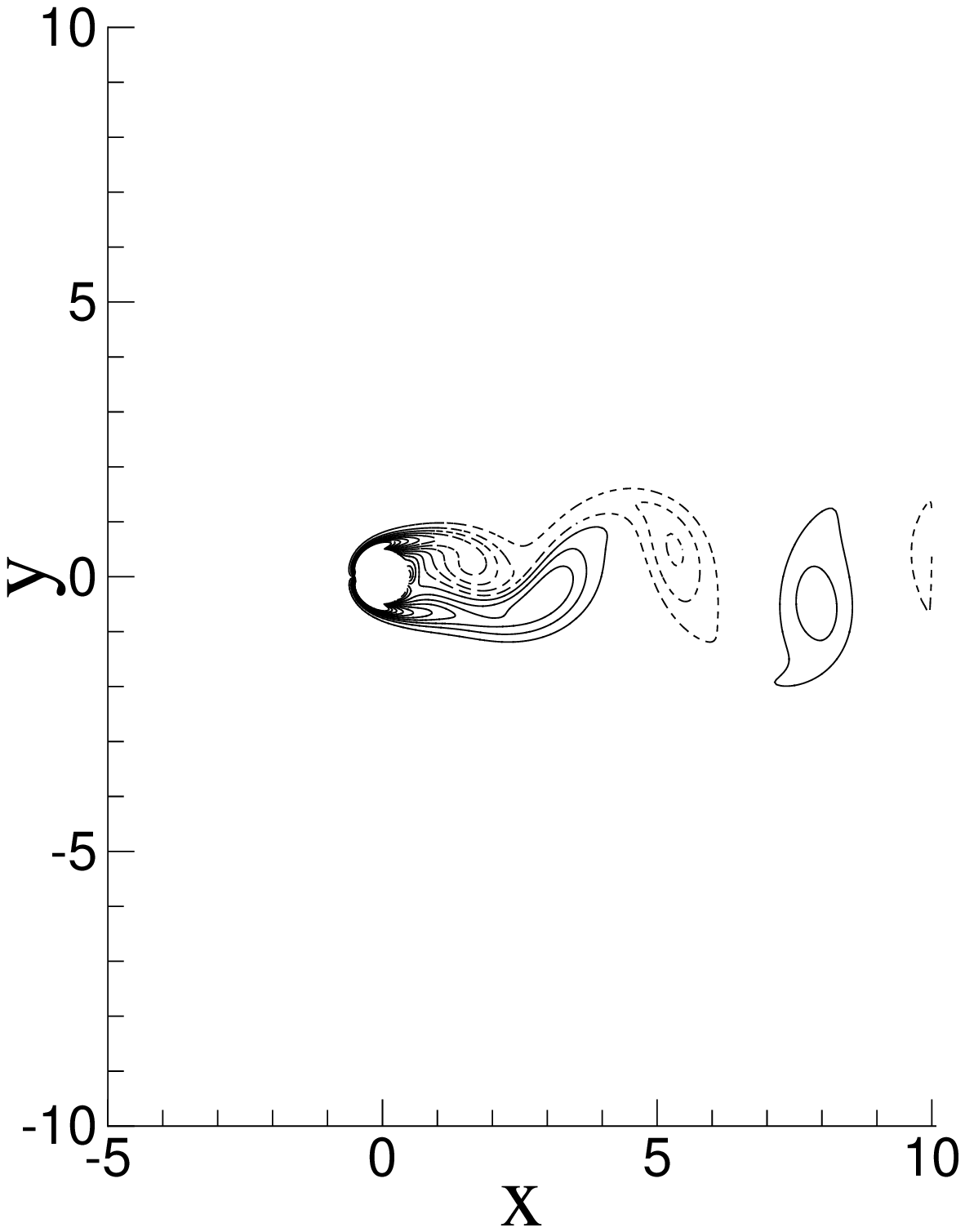}(b)
\includegraphics[width=2in]{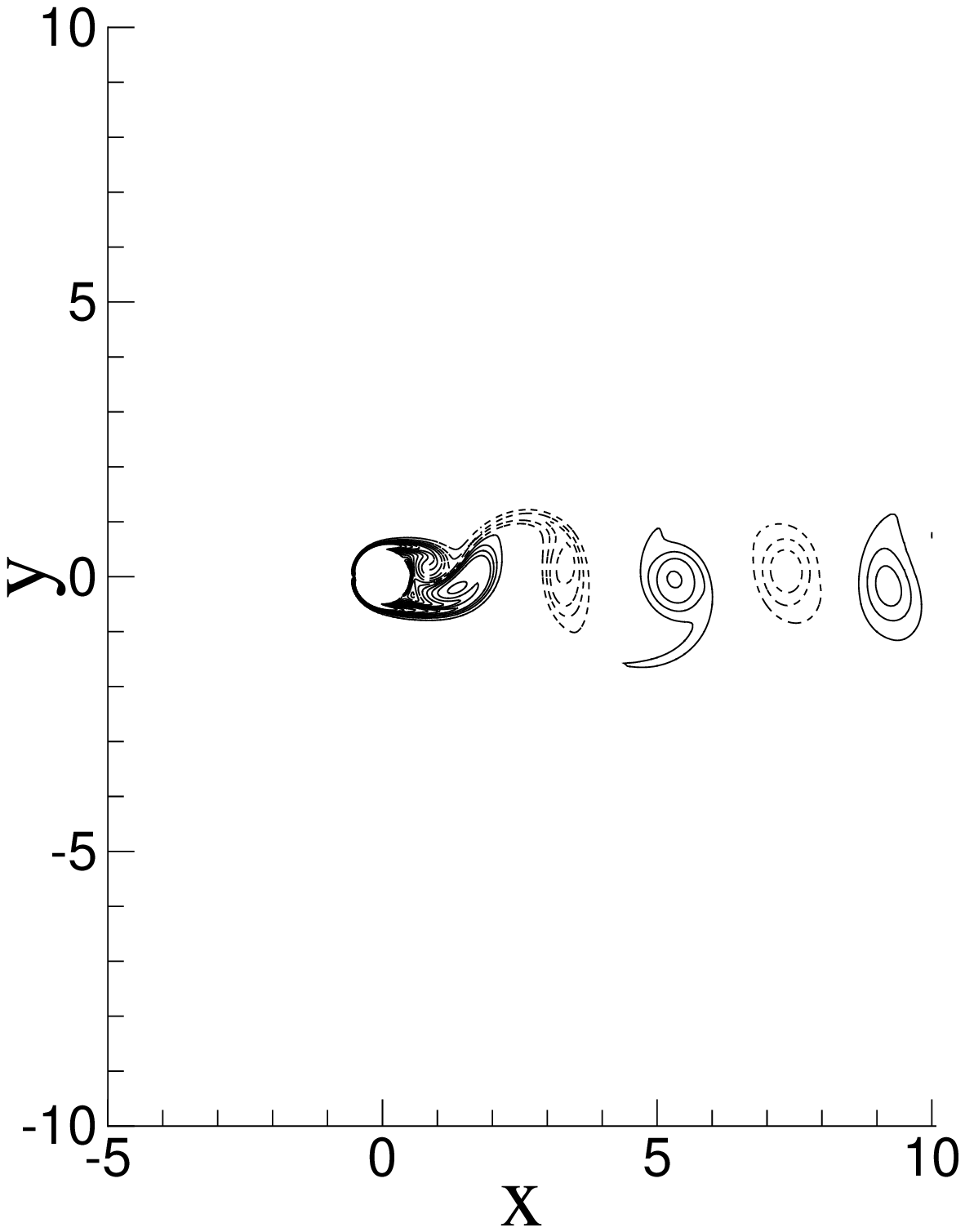}(c)
}
\caption{
Circular cylinder flow:
Contours of instantaneous vorticity at
Reynolds numbers 
(a) $Re=20$, 
(b) $Re=100$, and
(c) $Re=300$.
Dashed curves denote negative vorticity values.
}
\label{fig:cyl_vort}
\end{figure}

%

The general features of the circular cylinder
flow at Reynolds numbers of various flow regimes 
have been discussed in detail in \cite{Williamson1996}.
At Reynolds number around $Re=47$
the cylinder wake experiences an instability, 
and it becomes unsteady with vortex shedding from
the cylinder. The flow is  two-dimensional
at this point. 
When the Reynolds number increases to around $Re=180$,
another instability develops in the cylinder
wake, and the physical flow becomes three-dimensional.
In Figure \ref{fig:cyl_vort}
we show contours of the instantaneous
vorticity from our simulations at three
Reynolds numbers 
$Re=20$ (plot (a)), $100$ (plot (b)),
and $300$ (plot (c)).
These results correspond to the domain size $L=10D$ and
 the outflow condition OBC-E,
i.e. $(\theta,\alpha_1,\alpha_2)=(1,0,0)$
in \eqref{equ:bc_obc_gobc_1_reform}.
Figure \ref{fig:cyl_vort}(a)
corresponds to a steady-state flow,
while Figures \ref{fig:cyl_vort}(b) and (c)
show vortex shedding at the
higher Reynolds numbers.
These  are  two-dimensional
simulations.
In reality, the physical flow at $Re=300$
has already become three-dimensional.

\begin{figure}
\centerline{
\includegraphics[height=3in]{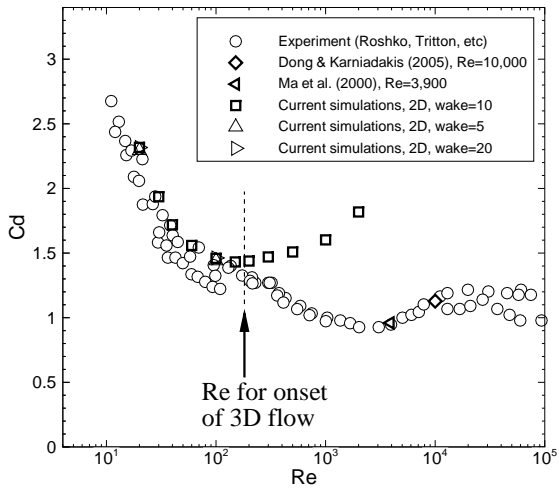}(a)
\includegraphics[height=3in]{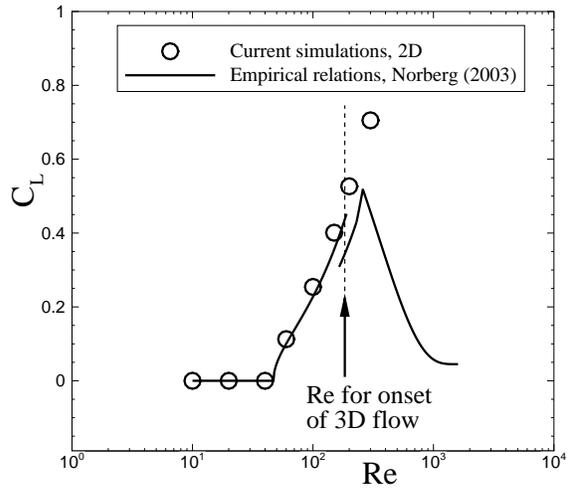}(b)
}
\caption{
Circular cylinder flow: Comparisons of (a) drag coefficients
and (b) RMS lift coefficients as a function of
the Reynolds number between current simulations
and experimental measurements.
Results of current simulations correspond to
the OBC-E outflow condition,
i.e. $(\theta,\alpha_1,\alpha_2)=(1,0,0)$.
}
\label{fig:Cd_CL_compare}
\end{figure}

We have computed the mean drag coefficient ($C_d$) and 
the root-mean-square (RMS) lift coefficient ($C_L$) 
on the cylinder from the simulations.
These coefficients are respectively defined as
\begin{equation}
C_d = \frac{\overline{F}_x}{\frac{1}{2}\rho U_0^2}, \qquad
C_L = \frac{F_y^{\prime}}{\frac{1}{2}\rho U_0^2},
\end{equation}
where
$\overline{F}_x$ is the mean (time-averaged) drag,
i.e. the $x$ component of force,
on the cylinder,
$F_y^{\prime}$ is the RMS of the lift,
and $\rho$ is the fluid density.
In Figure \ref{fig:Cd_CL_compare}(a)
we compare the mean drag coefficient as a function
of the Reynolds number between current
simulations and the experimental 
measurements of \cite{Wieselsberger1921,DelanyS1953,
Finn1953,Tritton1959,Roshko1961}.
The drag coefficients from 
the three-dimensional simulations
of \cite{MaKK2000,DongK2005} are also shown 
in the figure.
The results of the current simulations
are obtained using the OBC-E (i.e. 
$(\theta,\alpha_1,\alpha_2)=(1,0,0)$) open
boundary condition.
The majority are for 
the flow domain $L=10D$,
while at $Re=20$ and $Re=100$
results are also obtained using the domains
$L=5D$ and $L=20D$ for this group of
tests.
Note also that the current simulations are 
in 2-D.
One can observe that,
in the 2-D regime the drag coefficients
from current simulations are in good agreement
with the experimental data. 
In the 3-D regime, i.e. at Reynolds numbers
beyond about $Re=180$ when the physical
flow of the wake becomes three-dimensional,
one can observe a marked discrepancy 
between the drag coefficients from the current 2-D
simulations and the experimental data.
This discrepancy becomes more pronounced with
increasing Reynolds number.

Figure \ref{fig:Cd_CL_compare}(b) shows
a comparison of the RMS lift coefficient
as a function of the Reynolds number
between the current simulations
and the empirical relation given by 
Norberg \cite{Norberg2003}.
The simulation results are obtained
on the domain $L=10D$ with
the OBC-E open boundary condition.
In the 2-D regime,
the RMS lift results from current simulations
agree with the empirical relation
reasonably well. 
However, at Reynolds numbers in
the 3-D regime, the current simulations
significantly over-predict the
lift coefficient, which is a well-known
issue with 2-D simulations \cite{DongK2005,DongKER2006}.


\begin{table}
\begin{center}
\begin{tabular*}{0.7\textwidth}{@{\extracolsep{\fill}}
l c c}
\hline
Source & $Re=100$ & $Re=200$ \\
Braza et al. (1986) \cite{BrazaCH1986} & $0.21$ & $0.55$ \\
Karniadakis (1988) \cite{Karniadakis1988} & -- & $0.48$ \\
Engelman \& Jamnia (1990) \cite{EngelmanJ1990} & $0.26$ & -- \\
Meneghini \& Bearman (1993) \cite{MeneghiniB1993} & -- & $0.54$ \\
Beaudan \& Moin (1994) \cite{BeaudanM1994} & $0.24$ & -- \\
Zhang et al. (1995) \cite{Zhangetal1995} & $0.25$ & $0.53$ \\
Newman \& Karniadakis (1995) \cite{NewmanK1995} & -- & $0.51$ \\
Tang \& Audry (1997) \cite{TangA1997} & $0.21$ & $0.45$ \\
Persillon \& Braza (1998) \cite{PersillonB1998} & $0.27$ & $0.56$ \\
Zhang \& Dalton (1998) \cite{ZhangD1998} & -- & $0.48$ \\
Kravchenko et al. (1999) \cite{KravchenkoMS1999} & $0.22$ & -- \\
Hwang \& Lin (1992) \cite{HwangL1992} & $0.27$ & $0.42$ \\
Newman \& Karniadakis (1996) \cite{NewmanK1996} & $0.24$ & -- \\
Dong \& Shen (2010) \cite{DongS2010} & -- & $0.501$ \\
Franke et al. (1990) \cite{FrankeRS1990} & -- & $0.46$ \\
Current simulations (wake=5) & $0.261$ & -- \\
Current simulations (wake=10) & $0.254$ & $0.527$ \\
Current simulations (wake=15) & $0.253$ & -- \\
Current simulations (wake=20) & $0.253$ & -- \\
\hline
\end{tabular*}
\caption{ Circular cylinder flow: Comparison of 
RMS lift coefficients at $Re=100$ and $Re=200$
between current simulations and existing  simulations
from literature. 
}
\label{tab:circyl_lift_re100}
\end{center}
\end{table}

In Table \ref{tab:circyl_lift_re100} we list
the RMS lift coefficients at $Re=100$ and $Re=200$
from current simulations.
At $Re=100$ the lift coefficients have been obtained
on four domains $L=5D$, $10D$, $15D$ and $20D$,
while at $Re=200$ the result is for 
the domain $L=10D$.
For comparison, we have also listed in this table
the lift coefficients from existing
simulations from the literature
for these two Reynolds numbers. 
First, one can observe that the domain size (or the size of
the wake region) has a certain effect on
the lift coefficient. As the wake region 
increases to a certain size, e.g. about $L=10D$
at $Re=100$, the obtained lift coefficient
essentially will not change any longer or only change
very slightly.
Second, the lift coefficients from the existing simulations
in the literature exhibit a spread over a range of values.
The results from current simulations
appear in good agreement with the existing
simulation data, and lie well within the range of
existing data.


\begin{figure}
\centering
\includegraphics[width=4in]{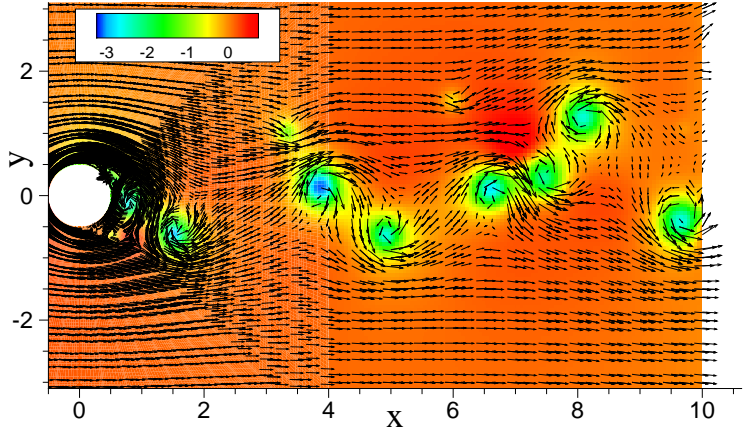}
\caption{ (color online)
Instantaneous velocity fields and pressure distributions (color contours)
of cylinder flow
at $Re=5000$, corresponding to the OBC-C outflow condition or
$(\theta,\alpha_1,\alpha_2)=(1,1,0)$.
Velocity vectors are plotted on every fifth quadrature point
in each direction within each element.
}
\label{fig:cyl_vel_pressure}
\end{figure}


%
%

Let us now focus on the stability issue with the outflow boundaries
at higher Reynolds numbers.
As the Reynolds number becomes large,
the vortices shed from the cylinder can 
persist a long time in the wake
before being sufficiently dissipated. 
For a given computational domain
with a certain size for the wake region,
as the Reynolds number becomes sufficiently large,
the strong vortices shed from
the cylinder will eventually reach
the outflow/open boundary.
These vortices can induce  backflows at
the open boundaries, and
with usual outflow/open boundary conditions
the simulations will instantly become
unstable. 
This is a well-known
numerical instability associated with
the open boundaries.

The open boundary conditions we presented
in Section \ref{sec:method} are effective
in dealing with this instability,
 because these conditions
ensure the energy stability of the system
even in the presence of strong vortices or
backflows at the open boundaries.
Figure \ref{fig:cyl_vel_pressure} shows
the instantaneous velocity fields
and the pressure distributions 
at Reynolds number 
$Re=5000$.
They are obtained with
the domain size $L=10D$.
%
One can clearly observe 
the strong vortices at
the open boundary at these Reynolds numbers.
The current open boundary conditions
and the pressure correction-based algorithm
 produce stable
simulations in these situations.
On the other hand,
we observe that with the traction-free
boundary condition (see e.g. \cite{SaniG1994})
or its variant the no-flux boundary condition
(i.e. $\frac{\partial\mathbf{u}}{\partial n}=0$
and $p=0$) the computation
blows up instantly when the vortices hit
the open boundary at these Reynolds numbers.
%


\begin{table}
\begin{center}
\begin{tabular*}{0.8\textwidth}{@{\extracolsep{\fill}}
l c c c}
\hline
$(\theta,\alpha_1,\alpha_2)$ or type & $Cd$ & $Cd_{rms}$ & $C_{L}$ \\
$(\frac{1}{2},0,0)$ or OBC-A & $1.743$ & $0.377$ & $1.440$ \\
$(1,0,1)$ or OBC-B & $1.738$  & $0.386$ & $1.434$ \\
$(1,1,0)$ or OBC-C & $1.738$ & $0.378$ & $1.428$ \\
$(0,1,0)$ or OBC-D & $1.736$ & $0.379$ & $1.423$ \\
$(1,0,0)$ or OBC-E & $1.726$ & $0.380$ & $1.420$ \\
$(0,0,0)$ or OBC-F & $1.699$ & $0.373$ & $1.403$ \\
\hline
\end{tabular*}
\caption{ Circular cylinder flow at $Re=5000$: 
Comparison of the mean drag coefficient $C_{d}$,
RMS drag coefficient $Cd_{rms}$ and 
RMS lift coefficient $C_L$ obtained with
different outflow boundary conditions
corresponding to different 
parameters $(\theta,\alpha_1,\alpha_2)$.
}
\label{tab:cyl_force_param_re5k}
\end{center}
\end{table}


Let us next consider the effect of different
open boundary conditions on the results.
We observe that the results obtained using the several
open boundary conditions from Section \ref{sec:obc} are quite similar.
In Table \ref{tab:cyl_force_param_re5k}
we have listed the mean drag coefficient $Cd$,
RMS drag coefficient $Cd_{rms}$, and
the RMS lift coefficient $C_L$
at $Re=5000$ obtained with the
several open boundary conditions in
Section \ref{sec:obc}.
The RMS drag coefficient is defined as
$
Cd_{rms} = \frac{F_x^{\prime}}{\frac{1}{2}\rho U_0^2},
$
where $F_x^{\prime}$ is the RMS of the drag.
One can observe that 
these force coefficients are quantitatively very 
close, with the maximum difference
on the order of $2\sim 3\%$.


\subsection{Impinging Jet on a Wall with Open Boundaries}
\label{sec:jet}



\begin{figure}
\centerline{
\includegraphics[height=3in]{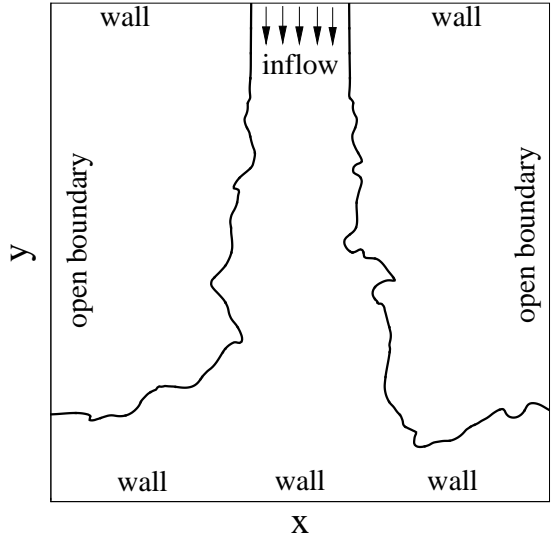}
}
\caption{
Flow configuration for the impinging jet on a wall with open boundaries.
}
\label{fig:jetcav_config}
\end{figure}

In this section we consider a jet impinging on a solid wall
involving open domain boundaries in two dimensions.
At moderate and high Reynolds numbers, the instability of
the jet and the presence of the open boundaries 
make this problem very challenging to simulate. 

We refer to Figure \ref{fig:jetcav_config}
for the configuration of this problem.
Consider a fluid jet, of diameter $D$, issuing
into a rectangular domain through the top wall. 
The domain has the following dimension,
$-\frac{5}{2}D \leqslant x\leqslant \frac{5}{2}D$
and $0\leqslant y\leqslant 5D$.
The top and bottom of the domain are solid walls,
while the left and right sides of the domain
are open, where the fluid can leave or enter the 
domain freely.
The center of the jet is aligned with 
the middle of the top wall.
we assume that at the inlet the jet velocity is along 
the vertical direction and has the following profile,
\begin{equation}
\left\{
\begin{split}
& 
u = 0 \\
&
v = -U_0\left[
  \tanh \frac{1-\frac{x}{R_0}}{\sqrt{2}\epsilon}\left[H(x,0) - H(x, R_0) \right]
  + \tanh \frac{1+\frac{x}{R_0}}{\sqrt{2}\epsilon}\left[H(x,-R_0) - H(x, 0) \right]
\right]
\end{split}
\right.
\label{equ:jet_prof}
\end{equation}
where $U_0=1$ is a velocity scale,
$R_0=\frac{D}{2}$ is the jet radius, and
$\epsilon=\frac{1}{40}D$.
$H(x,x_0)$ is the heaviside step function, taking unit value
if $x\geqslant x_0$ and vanishing otherwise.

All the length variables are normalized by the jet diameter $D$,
and all velocities are normalized by $U_0$.
The Reynolds number for this problem is defined by
equation \eqref{equ:Re_expr},
noting the specific physical meanings of $U_0$ and $D$
for this problem.
We assume that there is no external body force.


The domain has been discretized using  $400$ quadrilateral
elements of equal sizes,
with $20$ elements in both the $x$ and $y$ directions.
We impose the velocity Dirichlet boundary
condition \eqref{equ:bc_dbc} on the top and bottom
sides of the domain, where the boundary
velocity is set to $\mathbf{w}=0$ at the walls
and set according to equation \eqref{equ:jet_prof}
at the jet inlet.
On the left and right sides of the domain
the open boundary condition \eqref{equ:bc_obc_gobc_1_reform}
is imposed,
with $\mathbf{f}_b=0$ and $\delta = \frac{1}{100}$.
Different $(\theta,\alpha_1,\alpha_2)$ parameters
have been tested corresponding to
the open boundary conditions OBC-A to OBC-F.

We employ the algorithm developed in Section 
\ref{sec:method} in the simulations,
and have considered several Reynolds numbers
ranging from $Re=2000$ to $Re=10000$.
The element order in the simulations 
ranges from $12$ for $Re=2000$ 
to $16$ for $Re=10000$.
The time step size ranges from 
$\frac{U_0\Delta t}{D}=2.5 \times 10^{-4}$
for $Re=2000$
to $\frac{U_0\Delta t}{D} = 2 \times 10^{-4} $
for $Re=10000$.



\begin{figure}
\centerline{
\includegraphics[height=3in]{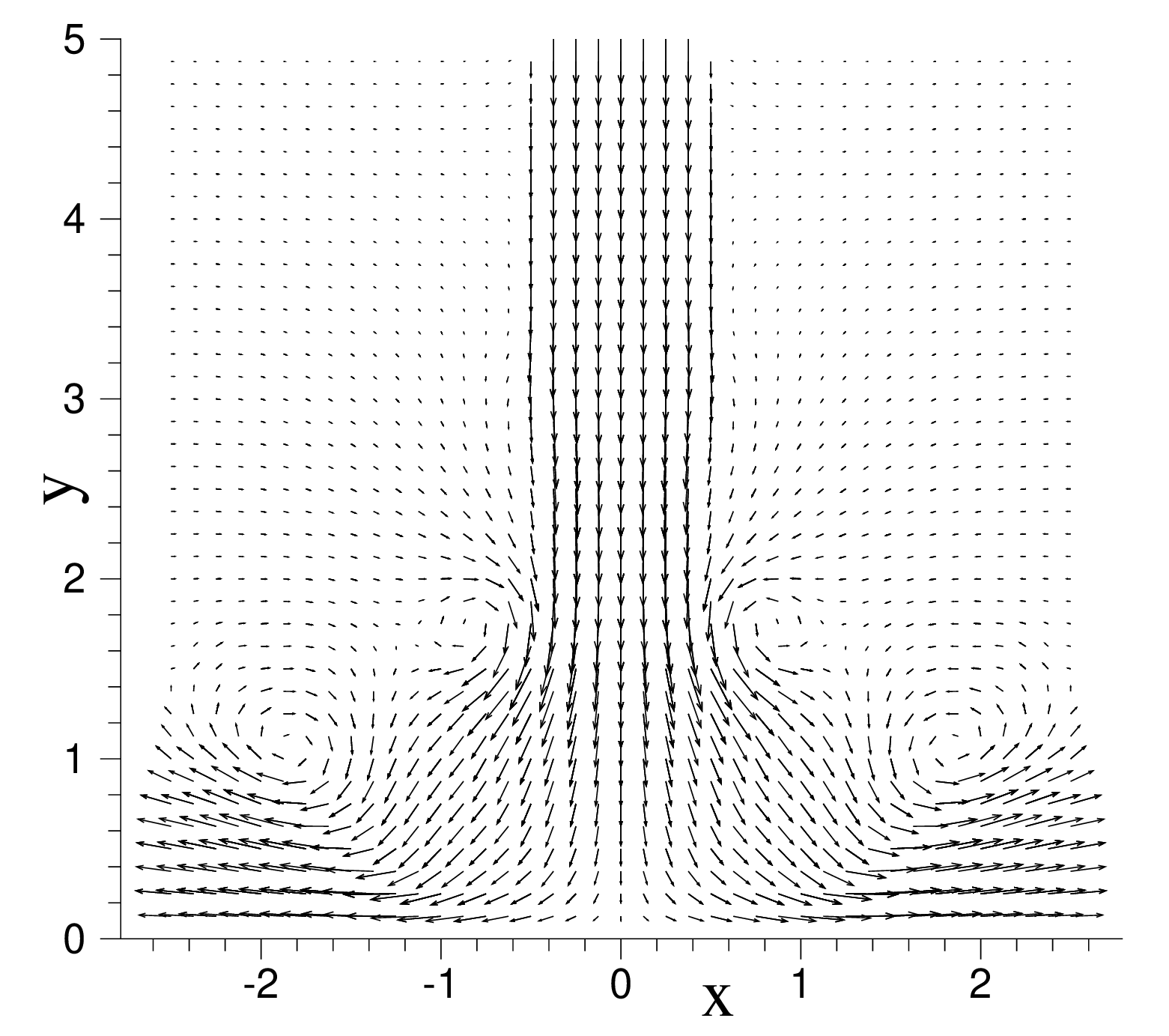}(a)
\includegraphics[height=3in]{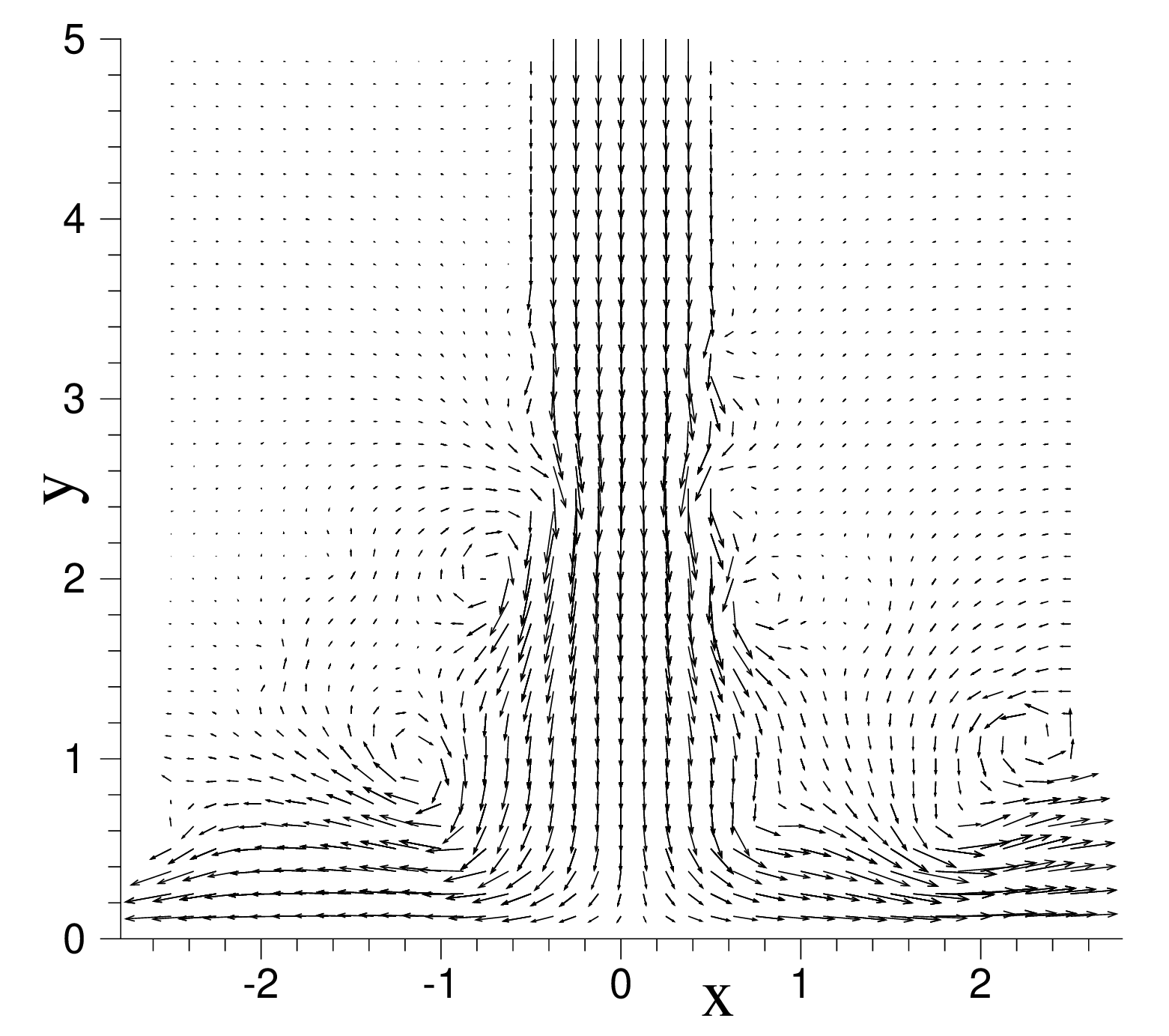}(b)
}
\caption{
Impinging jet on a wall:
Instantaneous velocity distributions at Reynolds numbers
(a) $Re=2000$ and 
(b) $Re=5000$,
obtained using the OBC-C open boundary condition
or $(\alpha,\alpha_1,\alpha_2)=(1,1,0)$.
Velocity vectors are plotted on every eleventh
quadrature point in each direction within each element.
}
\label{fig:jetcav_re2k_re5k}
\end{figure}


We first look into the basic features of
this flow. Figure \ref{fig:jetcav_re2k_re5k}(a)
shows the instantaneous velocity distribution
at $Re=2000$. The jet profile appears to be stable
within a certain distance downstream of
the inlet, $2\lesssim y/D\leqslant 5$
in this case.
Beyond this region, the instability causes
vortices to form along
the profile of the jet. 
After impinging onto the bottom wall
the jet splits into two streams,
which run out of the domain respectively through 
the left and the right open boundaries.
The vortices formed on the edges of the jet
are convected out of domain alongside
the horizontal streams.
The velocity distribution appears to be
symmetric about the jet centerline ($x=0$)
at this Reynolds number.

Figure \ref{fig:jetcav_re2k_re5k}(b)
shows the distribution of the instantaneous
velocity at a higher Reynolds number
$Re=5000$. One can observe a 
 basic feature similar to that of $Re=2000$.
However, the region with a stable jet
profile downstream of the inlet is smaller,
with $3.5\lesssim y/D\leqslant 5$
at $Re=5000$.
In addition, the velocity distribution
has lost the symmetry about
the jet centerline.


The vortices formed along the jet profile
cause the backflow instability  on
the left and right open boundaries,
and pose a severe challenge to the
simulation of this problem.
The open boundary conditions and
the numerical algorithm developed
in Section \ref{sec:method}
are crucial to the stability of
the simulations.
In contrast, usual open boundary
conditions such as the 
traction-free condition and
the no-flux condition
are unstable for the Reynolds numbers
simulated here, and
we observe that the computation
blows up instantly when the vortices
hit the open boundaries.

\begin{table}
\begin{center}
\begin{tabular*}{0.6\textwidth}{@{\extracolsep{\fill}}l c c}
\hline 
$(\theta,\alpha_1,\alpha_2)$ or type & Mean $F_y$ & RMS $F_y$ \\
$(\frac{1}{2},0,0)$ or OBC-A & $-1.015$ & $0.241$ \\
$(1,0,1)$ or OBC-B & $-1.005$ & $0.181$ \\
$(1,1,0)$ or OBC-C & $-1.009$ & $0.220$ \\
$(0,1,0)$ or OBC-D & $-1.016$ & $0.231$ \\
$(1,0,0)$ or OBC-E & $-0.984$ & $0.156$ \\
$(0,0,0)$ or OBC-F & $-1.024$ & $0.258$ \\
\hline
\end{tabular*}
\end{center}
\caption{
Impinging jet on a wall at $Re=10000$:
Mean and RMS forces (vertical component) on the wall
obtained using different open  boundary conditions.
}
\label{tab:jetcav_re10k_force_param}
\end{table}

We have performed long-time simulations of
this flow using different open boundary conditions.
%
A quantitative comparison is shown in Table
\ref{tab:jetcav_re10k_force_param},
where we have listed the time-averaged mean
and the RMS forces (vertical component)
on the wall obtained using
the several open boundary conditions 
from Section \ref{sec:obc}.
The mean forces corresponding to
different  boundary conditions
are very close, with
the maximum difference among them
about $4\%$.
The RMS forces obtained with different
open boundary conditions are
also comparable.
While the RMS forces with the other
open boundary conditions are quite close,
those corresponding to 
OBC-E and OBC-B are somewhat smaller.



\section{
A Provably Unconditionally Stable Scheme  for
a Sub-Class of Open Boundary Conditions}

In this section we briefly discuss a rotational pressure correction scheme
with a provable unconditional stability for the following sub-class of 
the open boundary conditions \eqref{equ:bc_obc_gobc_1},
\begin{equation}
\theta = \alpha_2 = 0, \quad
\alpha_1 \geqslant 0.
\end{equation}
The boundary conditions OBC-D and OBC-F belong to
this sub-class.


Our discussions here will be limited to the temporal discretization only,
and it is assumed that the field variables are continuous in space
in this section. The stability proof is provided for the scheme
with a nominal first order in time. 
We assume a homogeneous velocity Dirichlet condition
on $\partial\Omega_d$, i.e. $\mathbf{w}=0$ in \eqref{equ:bc_dbc},
and that there is no external body force, i.e. $\mathbf{f}=0$
in \eqref{equ:nse}.
In addition, we assume that $\mathbf{f}_b=0$ in the open boundary
condition \eqref{equ:bc_obc_gobc_1}, and that $\delta \rightarrow 0$
in the $\Theta_0(\mathbf{n},\mathbf{u})$ function, that is,
\begin{equation}
\lim_{\delta\rightarrow 0}\Theta_0(\mathbf{n},\mathbf{u}) = \Theta_{s0}(\mathbf{n},\mathbf{u})
= \left\{\begin{array}{ll}
1, & \text{if} \ \mathbf{n}\cdot\mathbf{u}<0, \\
0, & \text{otherwise}.
\end{array}
\right.
\end{equation}


Given $(\tilde{\mathbf{u}}^n,\mathbf{u}^n,p^n)$, we compute
these field variables at time step $(n+1)$ as follows.
First, find $\tilde{\mathbf{u}}^{n+1}$ satisfying
\begin{subequations}
\begin{equation}
\frac{\tilde{\mathbf{u}}^{n+1} - {\mathbf{u^n}}}{\Delta t}
+ \mathbf{u^n}\cdot \nabla \tilde{\mathbf{u}}^{n+1}
+\nabla p^n - \nu\nabla^2\tilde{\mathbf{u}}^{n+1}
= 0,
\label{equ:v1_A}
\end{equation}
\begin{equation}
\tilde{\mathbf{u}}^{n+1} = 0,
\qquad \text{on} \ \partial\Omega_d,
\label{equ:v2_A}
\end{equation}
\begin{equation}
- p^{n}\mathbf{n}
+\nu \mathbf{n}\cdot\nabla\tilde{\mathbf{u}}^{n+1}
   -(1+\alpha_1)\frac 12 (\mathbf{u}^n\cdot \mathbf{n}) \tilde{\mathbf{u}}^{n+1}\Theta_{s0}(\mathbf{n},\mathbf{u}^{n})=0,
\qquad \text{on} \ \partial\Omega_o.
\label{equ:v3_A}
\end{equation}
\end{subequations}
Then, find $(\mathbf{u}^{n+1},p^{n+1})$ satisfying
\begin{subequations}
\begin{equation}
\frac{\mathbf{u}^{n+1}-\tilde{\mathbf{u}}^{n+1}}{\Delta t}
+ \nabla\left(p^{n+1} - p^n + \chi\nabla\cdot \tilde{\mathbf{u}}^{n+1} \right) = 0,
\qquad \label{equ:p1_A}
\end{equation}
\begin{equation}
\nabla\cdot\mathbf{u}^{n+1} = 0,
\label{equ:p2_A}
\end{equation}
\begin{equation}
\mathbf{n}\cdot\mathbf{u}^{n+1} = 0,
\qquad \text{on} \ \partial\Omega_d,
\label{equ:p3_A}
\end{equation}
\begin{equation}
p^{n+1} - p^n + \chi\nabla\cdot \tilde{\mathbf{u}}^{n+1}=0, 
\qquad \text{on} \ \partial\Omega_o,
\label{equ:p4_A}
\end{equation}
\end{subequations}
where $\chi$ is a positive constant to be specified subsequently.

Note that equations \eqref{equ:p1_A}--\eqref{equ:p4_A} are equivalent to the following,
\begin{subequations}
\begin{equation}
\nabla^2\left(p^{n+1} - p^n + \chi\nabla\cdot \tilde{\mathbf{u}}^{n+1} \right) = \frac{\nabla\cdot\tilde{\mathbf{u}}^{n+1}}{\Delta t},
\qquad \label{equ:p1B}
\end{equation}
\begin{equation}
\frac{\partial}{\partial \mathbf{n}}( p^{n+1} - p^n + \chi\nabla\cdot \tilde{\mathbf{u}}^{n+1}) =0 \quad \text{on} \ \partial\Omega_d;
\qquad 
p^{n+1} - p^n + \chi\nabla\cdot \tilde{\mathbf{u}}^{n+1}=0 \quad \text{on} \ \partial\Omega_o,
\label{equ:p2B}
\end{equation}
\end{subequations}
and
\begin{equation}\label{equ:p3B}
\mathbf{u}^{n+1} = \tilde{\mathbf{u}}^{n+1} 
- \Delta t\nabla(p^{n+1}-p^n +\chi\nabla\cdot\tilde{\mathbf{u}}^{n+1}).
\end{equation}

To prove the stability of the scheme given by \eqref{equ:v1_A}--\eqref{equ:p4_A},
we define the auxiliary variables $q^n$ and $\psi^n$ by
  $q^0=-p^0$, 
and 
\begin{equation}\label{defpsi}
q^{n+1}=q^n+\chi\nabla\cdot \tilde{\mathbf{u}}^{n+1},\quad \psi^{n+1}=p^{n+1}+q^{n+1}.
\end{equation}
Then, equations \eqref{equ:p1_A} and \eqref{equ:p4_A} can be written as:
%
\begin{equation}
\frac{\mathbf{u}^{n+1}-\tilde{\mathbf{u}}^{n+1}}{\Delta t}
+ \nabla\left(\psi^{n+1} - \psi^n \right) = 0,
\qquad \label{equ:p1_B}
\end{equation}
%
%
%
\begin{equation}
\psi^{n+1} - \psi^n =0, \qquad \text{on} \ \partial\Omega_o.
\label{equ:p4_B}
\end{equation}
%
Note that $\psi^0=p^0+q^0=0$, and hence we have 
\begin{equation}
\psi^n=0, \quad 
\text{on} \ \partial\Omega_o,
\label{equ:psi_n_outflow}
\end{equation}
for all $n$
based on equation \eqref{equ:p4_B}.

Let $(f,g)$ denote the $L^2$ inner product between field variables
$f(\mathbf{x},t)$ and $g(\mathbf{x},t)$, and define 
$\|f\|^2 = (f,f)$.
Taking the $L^2$ inner 
product between \eqref{equ:v1_A} 
and $2\Delta t \tilde{\mathbf{u}}^{n+1}$, and noticing that
(since $\nabla \cdot \mathbf{u^n}=0$)
\begin{equation*}
(\mathbf{u^n}\cdot \nabla\tilde{\mathbf{u}}^{n+1},\tilde{\mathbf{u}}^{n+1})=\frac 12 \int_{\partial\Omega_o} (\mathbf{u}^n\cdot \mathbf{n}) |\tilde{\mathbf{u}}^{n+1}|^2,
\end{equation*}
 we obtain
\begin{equation}
\begin{split}
 \|\tilde{\mathbf{u}}^{n+1}\|^2&-\|\mathbf{u}^n\|^2+\|\tilde{\mathbf{u}}^{n+1}-\mathbf{u}^n\|^2+2\nu\Delta t\|\nabla \tilde{\mathbf{u}}^{n+1}\|^2
 -2\Delta t(p^n, \nabla \cdot \tilde{\mathbf{u}}^{n+1})\\
 &
 =2\Delta t \int_{\partial{\Omega_o}} (\nu \mathbf{n}\cdot\nabla\tilde{\mathbf{u}}^{n+1}-p^n \mathbf{n}) \cdot \tilde{\mathbf{u}}^{n+1}
 -2\Delta t (\mathbf{u^n}\cdot \nabla\tilde{\mathbf{u}}^{n+1},\tilde{\mathbf{u}}^{n+1})\\
 &
 = 2\Delta t \int_{\partial{\Omega_o}}  \left(
 	  \nu \mathbf{n}\cdot\nabla\tilde{\mathbf{u}}^{n+1}-p^n \mathbf{n} 
	  - \frac{1}{2} (\mathbf{u}^n\cdot \mathbf{n}) \tilde{\mathbf{u}}^{n+1}
 	\right) \cdot \tilde{\mathbf{u}}^{n+1} \\
 &
 =\Delta t \int_{\partial\Omega_o} \left[
 		(\mathbf{u}^n\cdot \mathbf{n}) |\tilde{\mathbf{u}}^{n+1}|^2 \left( \Theta_{s0}(\mathbf{n}, \mathbf{u}^{n})-1 \right) 
		+ \alpha_1 (\mathbf{u}^n\cdot \mathbf{n}) |\tilde{\mathbf{u}}^{n+1}|^2 \Theta_{s0}(\mathbf{n}, \mathbf{u}^{n})
	\right] \\
 &
 \leqslant 0,
 \end{split}
\end{equation}
where we have used integration by part, the divergence theorem,
and the equation \eqref{equ:v3_A}.

We deal with the term 
$-2\Delta t(p^n, \nabla \cdot \tilde{\mathbf{u}}^{n+1})=-2\Delta t(\psi^n-q^n, \nabla \cdot \tilde{\mathbf{u}}^{n+1})$ as follows.
Note that
\begin{equation}
 \begin{split}
 -2\Delta t(\psi^n, \nabla \cdot \tilde{\mathbf{u}}^{n+1})&=
 2\Delta t(\nabla \psi^n, \tilde{\mathbf{u}}^{n+1})
 =2\Delta t(\nabla\psi^n, \mathbf{u}^{n+1}+\Delta t\nabla (\psi^{n+1}-\psi^n))\\
&= 2\Delta t^2(\nabla \psi^n, \nabla (\psi^{n+1}-\psi^n))\\
 &=\Delta t^2(\|\nabla \psi^{n+1}\|^2-\|\nabla \psi^{n}\|^2-
 \|\nabla (\psi^{n+1}-\psi^n)\|^2)\\
 &=\Delta t^2(\|\nabla \psi^{n+1}\|^2-\|\nabla \psi^{n}\|^2)-\|\mathbf{u}^{n+1}-\tilde{\mathbf{u}}^{n+1}\|^2,
 \end{split}
\end{equation}
where we have use \eqref{equ:p2_A}, \eqref{equ:p1_B},
and \eqref{equ:psi_n_outflow}.
Note also that
\begin{equation}
 \begin{split}
 2\Delta t(q^n, \nabla \cdot \tilde{\mathbf{u}}^{n+1}) 
 &=\frac{2\Delta t}{\chi} (q^n,q^{n+1}-q^n) \\
 &
 =\frac{\Delta t}{\chi} \left(\|q^{n+1}\|^2 - \|q^{n}\|^2- \|q^{n+1}-q^{n}\|^2 \right) \\
 &=\frac{\Delta t}{\chi} (\|q^{n+1}\|^2 - \|q^{n}\|^2) - \chi\Delta t \|\nabla \cdot \tilde{\mathbf{u}}^{n+1}\|^2.
\end{split}
\end{equation}
Next we take the $L^2$ inner product between \eqref{equ:p1_B} and $2
\mathbf{u}^{n+1}$ to obtain
\begin{equation}
\|\mathbf{u}^{n+1}\|^2 -\|\tilde{\mathbf{u}}^{n+1}\|^2
+\|\mathbf{u}^{n+1}-\tilde{\mathbf{u}}^{n+1}\|^2=0.
\end{equation}
Combining the above four relations leads to
\begin{equation}
  \|E^{n+1}\|^2+\|\tilde{\mathbf{u}}^{n+1}-\mathbf{u}^n\|^2
    +\Delta t\left(2\nu\|\nabla \tilde{\mathbf{u}}^{n+1}\|^2-\chi\|\nabla \cdot \tilde{\mathbf{u}}^{n+1}\|^2 \right)
\leqslant \|E^n\|^2,
\end{equation}
where 
\begin{equation*}
\|E^{k}\|^2:=\|\mathbf{u}^k\|^2+\Delta t^2\|\nabla \psi^{k}\|^2+ \frac{\Delta t}{\chi}\|q^{k}\|^2.
\end{equation*}
We recall that
\begin{equation}
 \|\nabla \cdot \mathbf{u}\|^2\leqslant d \|\nabla  \mathbf{u}\|^2,
 \quad \forall u\in H^1(\Omega)^d,
\end{equation}
where $d$ is the dimension of the domain.
Hence, we can conclude from the above that 

\noindent\underline{\bf Theorem 1.}
Let $0<\chi\leqslant \frac {2\nu}d$. Then the scheme \eqref{equ:v1_A}-\eqref{equ:p4_A}
is unconditionally stable, and we have
\begin{equation*}
  \|\mathbf{u}^{n+1}\|^2+\Delta t^2\|\nabla \psi^{n+1}\|^2+\frac{\Delta t}{\chi}\|q^{n+1}\|^2
    +\Delta t(2\nu-\chi d)\|\nabla \tilde{\mathbf{u}}^{n+1}\|^2
  \leqslant \|\mathbf{u}^n\|^2+\Delta t^2\|\nabla \psi^{n}\|^2+\frac{\Delta t}{\chi}\|q^{n}\|^2.
\end{equation*}

\section{Concluding Remarks}
\label{sec:summary}

%
%
%
%

We have presented a generalized
form of open/outflow boundary conditions for incompressible flows 
and a pressure correction-based algorithm for numerically
treating these open boundary conditions.
The generalized form represents a family of open
boundary conditions, 
with the characteristic that they all
ensure the energy stability of the system. 
These open boundary conditions are effective 
even when strong backflows or vortices occur
at the open/outflow boundaries.
Our algorithm is based on a 
rotational pressure-correction strategy, and
introduces
an auxiliary variable and an associated discrete
equation together with boundary conditions.
The formulation allows
for the direct computation of the pressure 
in the $H^1(\Omega)$ space. The algorithm imposes 
on the open boundary 
a pressure Dirichlet type condition in the pressure substep
and a velocity Neumann type condition
in the velocity substep.
The current algorithm can work properly with
equal orders of approximation
for the pressure and the velocity.

In addition to the above algorithm, which is
semi-implicit and conditionally stable in nature,
for a sub-class of the generalized form of open
boundary conditions we have also developed
an unconditionally stable scheme 
and provided a proof for its unconditional
stability.

Extensive numerical experiments have been presented
for several problems involving outflow/open
boundaries for a range of Reynolds numbers.
We have compared the current simulation results
with the experimental data from the literature,
as well as with the existing numerical simulations
by other researchers, 
to demonstrate the accuracy of the 
method developed in this work.
We have also shown that
our method produces long-time stable simulations
at moderate and high Reynolds numbers 
when strong vortices and backflows
occur at the open/outflow boundaries.
By contrast, usual outflow boundary conditions
such as the traction-free condition or
the no-flux condition encounter numerical difficulties
at these Reynolds numbers, and
the computation blows up instantly 
when the vortices hit the open boundary.


The numerical instability associated with
strong vortices or backflows at
the open/outflow boundaries 
are widely encountered in 
flow problems involving physically
unbounded domains.
The method developed in the current work
provides an effective means for overcoming 
this instability. 
It provides the opportunity for 
using a  substantially smaller
computational domain in numerical
simulations than otherwise 
for problems on physically unbounded
domains. The domain size
can be chosen solely based on the
consideration of physical accuracy.
The ability to use
a substantially smaller
computational domain will 
facilitate simulations at 
high Reynolds numbers,
because of the increased grid 
resolution under identical grid sizes.
The current method will be instrumental in
 numerical simulations 
at Reynolds numbers significantly
higher than the state of the art.


\section*{Acknowledgement}
S.D. would like to acknowledge the support 
from NSF (DMS-1318820) and 
ONR (N000141110028). J.S. would like to
acknowledge the support from NSF (DMS-1217066 and DMS-1419053).



%
\bibliographystyle{plain}
\bibliography{obc,mypub,nse,sem,cyl}

\end{document}